\documentclass{pas}

\usepackage{amsmath}
\usepackage{mathrsfs}
\usepackage{aas_macros}
\usepackage{hyperref} 
\hypersetup{colorlinks,citecolor=blue,linkcolor=blue,urlcolor=blue}
\usepackage{caption}
\usepackage{adjustbox}
\usepackage{tabularx}
\usepackage{comment}
\usepackage{multirow}

\newcommand\hi{\mbox{\sc Hi}}
\newcommand\ari{\mbox{\sc Ar~i}}
\newcommand\oi{\mbox{\sc O~i}}

\newcommand\kms{km s$^{-1}$}

\begin{document}

\graphicspath{{./}{figures/}}
\lefttitle{Publications of the Astronomical Society of Australia}
\righttitle{Aishani Das-Ghosh}

\jnlPage{1}{4}
\jnlDoiYr{2021}
\doival{10.1017/pasa.xxxx.xx}

\articletitt{Research Paper}

\title{Probing the Properties of Neutral Clouds in the Vicinity of the Local Bubble: Linking three-dimensional Dust with Neutral Hydrogen Absorption}

\author{\sn{Aishani} \gn{Das-Ghosh}$^{1}$, \sn{Nickolas} \gn{Pingel}$^{2, 3}$, and \sn{Snežana} \gn{Stanimirović}$^{3}$ }

\affil{$^1$Department of Engineering Sciences and Applied Mathematics, Northwestern University, 2145 Sheridan Rd, Evanston, IL 60201}
\affil{$^2$ Department of Astronomy, Indiana University, 727 East Third Street, Bloomington, IN 47405, USA} 
\affil{$^3$ Department of Astronomy, University of Wisconsin–Madison, 475 N Charter St, Madison, WI 53703, USA}

\corresp{N. Pingel, Email: nmpingel@iu.edu}


\history{(Received 24 01 2026; revised 01 08 2026; accepted 16 08 2026)}

\begin{abstract}
The Local Bubble (LB) contains significant internal structure, with many interstellar clouds lying within or on its walls. By observing time variability of atomic neutral hydrogen (\hi) absorption profiles in the direction of pulsars, it was suggested that some of the \hi\ structures exhibit variations in optical depth representative of the tiny scale atomic structure (TSAS). However, the exact location of these structures has not been known and understanding their origin in the framework of LB formation and evolution has been impossible. We use three-dimensional dust maps to extract density profiles of the hydrogen volume density along the line-of-sight (LOS) to four selected pulsars. We attempt to correlate individual density components with the Gaussian components of the \hi\ absorption spectra, revealing the TSAS towards B1929+10 likely lies outside of the LB. Comparing column density images integrated over the entire LOS and over sub-volumes of the 3D dust maps show a mixture of diffuse and clumpy structure. Using estimates of the mean \hi\ volume density and maximum total neutral hydrogen density along the LOS, we place upper and lower limits on ionisation fraction and thermal pressure, respectively, to characterise the properties of the local ISM towards each pulsar. Our method of using \hi\ absorption data in conjunction with three-dimensional dust data holds potential for broader applications, especially to validate the structures and properties of dust clouds in the LB and beyond.
\end{abstract}

\begin{keywords}
Our galaxy $-$ Interstellar clouds $-$ interstellar medium $-$ interstellar dust
\end{keywords}

\maketitle
\section{Introduction}
The local diffuse interstellar medium (ISM), consisting mostly of gas and dust, is characterised by ionisation state, temperature, and density. The multiphase model of the ISM establishes that the neutral ISM consists of two distinct phases, the warm neutral medium (WNM) and the cold neutral medium (CNM), which are embedded in the hot low-density medium (HIM; \citealt{field1969, McKee1977, wolfire2003}). Traditionally, the CNM and WNM are understood as being two thermal equilibrium states of the neutral medium. The theoretically expected properties of the CNM and WNM, based on the heating and cooling balance, are: a kinetic temperature $T_{k}\sim60-260$ K and a volume density of $n_{H}\sim7-70$ cm$^{-3}$ for the CNM , and $T_{k} = 5000-8300$ K and $n_{H}\sim0.2-0.9$ cm$^{-3}$ for the WNM (\citealt{wolfire2003}, Table 3 for the Solar neighbourhood). The WNM and CNM have been observed via the 21-cm (\hi) emission and absorption (e.g., \citealt{dickey2003, heiles2003b, fukui2014}). The diffuse ISM phases (neutral and ionised) have a roughly comparable thermal pressure \citep{ferriere1998}. 

The Local Bubble (LB) has been widely studied for its interesting properties, such as its irregular shape, porosity, and hydrodynamic instabilities on the surface of the cavity walls \citep{stanimirovic2010, welsh2010, oneill2024}. Thought to be the result of several supernova explosions taking place over the past 10-15 Myrs, the LB has pushed out the evacuated matter into a dense shell of gas and dust. Early investigation of the properties of \hi\ in the LB revealed the existence of diffuse clouds ($n_{H} = 0.01$ cm $^{-3}$) with high temperatures ($\sim10^{4}$ K) within the LB \citep{cox1987}. 

While warm, low-density clouds have been extensively studied inside the LB
\citep{frisch2011}, cold dense \hi\ clouds have been found widespread in the Solar neighbourhood (e.g., \citealt{begum2010}) but their exact location in relation to the LB has been hard to constrain. The Local Leo Cloud is a spectacular example of a cold \hi\ cloud located 45 pc from the Sun, with a mean temperature of 20 K, mean \hi\ column density of $2.5 \times 10^{19}$ cm $^{-2}$, and an astounding\ thermal gas pressure of $\sim6\times10^4$ cm$^{-3}$ K \citep{meyer2006, peek2011, meyer2012}. The combination of cold temperature, high pressure, and diffuse structure raises questions on the formation of such interstellar clouds in the seemingly (mostly) hot LB. Assuming that the cloud is in thermal equilibrium, the temperature and column density imply an extremely thin ($< 0.1$ pc), sheetlike geometry \citep{meyer2012}. 

A recent comprehensive analysis by \cite{zucker2022} using 3D dust data from \cite{leike2020} provided a major improvement in disentangling  structure and physical properties of interstellar clouds in the vicinity of the LB. In particular, the authors found that all major molecular clouds $\sim200$ pc from the Sun lie on the surface of the LB, and most clouds associated with the LB wall show either sheet-like or filamentary morphology. In modelling the expansion of the LB as a function of time, \cite{zucker2022} estimated the density of the ambient interstellar medium before the first supernova to be $n_{0} = 2.7^{+1.57}_{-1.02}$ cm$^{-3}$, indicating a mixture of the CNM and WNM. Thus, its overall formation and current structure of walls of the LB make for interesting interactions between different temperatures and phases of ISM that lead to the formation of molecular clouds.

With interactions between WNM and CNM and the hot diffuse gas inside the LB leading to a ``survival of the fittest", tiny-scale atomic structure (TSAS)--CNM structures on AU spatial scales--is a phenomenon that may potentially result from fragmentation and instabilities on the walls of the LB. \citet{stanimirovic2010} investigated \hi\ absorption towards 5 pulsars with the aim of capturing signatures of TSAS that manifest as fluctuations in the HI optical depth spectra. In the direction of one pulsar, B1929+10, the authors found repeated changes in the depth of the \hi\ optical depth, suggesting the existence of cold \hi\ structures on spatial scales $\sim5$-45 AU with estimated temperature of $\sim$50$-$200 K and inferred density of few $\times$ 10$^4$ cm$^{-3}$. However, since \hi\ absorption spectra provide only a one-dimensional probe,  understanding the origin, morphology, and exact spatial position of these absorbing structures was impossible until recent advances in relating dust extinction to three dimensional (3D) structures. Also, while integrated quantities like column density can be measured from spectra, volume density calculations require an assumption on the line-of-sight structure/cloud length. Structures elongated along the line-of-sight would result in a much lower volume densities and thermal pressure (e.g. \citep{heiles1997}). Therefore, observational constraints of the physical location and extent of TSAS is of great importance for understanding the evolution of the LB.  

The advent of large stellar surveys such as 2MASS \citep{skrutskie2006}, PanSTARRS \citep{kaiser2002} and SDSS/APOGEE \citep{albareti2017} and WISE \citep{wright2010} paved the way for several reconstructions of the local three dimensional (3D) dust distribution within $\sim$0.5 kpc from the Sun that may allow for greater analysis of the structure and properties of the neutral hydrogen under the assumption that dust and neutral gas are well mixed throughout the ISM. They utilise photometric measurements, and spectra for thousands of stars, from which the calculation of photometric distances is possible. For example, \cite{green2018} created a 3D dust map by combining the star data of Pan-STARRS and 2MASS, binning it in angular and distance bins, and performing independent Bayesian reconstructions per angular bin. The resulting dust map covers approximately three-quarters of the sky out to distances of $\sim$2 kpc. Building on the methodology of \citet{leike2020}, \citet{edenhofer24} constructed an expanded three-dimensional dust map using the updated extinction catalogue of \citet{zhang2023}. The substantially improved distance and extinction estimates for more than 220 million stars, combined with the use of Metric Gaussian Variational Inference to efficiently optimise the high-dimensional dust density model, enabled a significant increase in both the spatial resolution and fidelity of the reconstructed dust distribution. Based on the map from \citet{edenhofer24}, \citet{oneill2024} create an improved model of the LB accurate to the irregular structure and densities of the LB walls.

Due to the higher densities ($n\sim$10$-$100 cm$^{-3}$) of the CNM phase of \hi, collisions are sufficient to completely thermalise the \hi\ 21 cm hyperfine transition, meaning that the spin temperature ($T_{s}$)---measured unambiguously from \hi~absorption detections and describes the relative level population between the two hyperfine levels---is equal to the kinetic temperature ($T_k$). We utilise the 3D representation of the total neutral hydrogen (atomic+molecular) distribution inferred from the 3D dust maps of \citet{edenhofer24} in conjunction with the temperatures provided by the \hi~absorption measurements of \citet{stanimirovic2010} to probe local ISM properties. Specifically, we use the density and temperature information towards four of the five pulsars: B0823+26, B1133+16, B1929+10, and B2016+28, as they fall within the captured distance range (1.25 kpc) of the \citep{edenhofer24} map. The primary goals of this project are to localise the absorbing CNM structures seen against these pulsars to investigate their relation to the LB and provide independent constraints on the thermal pressure and ionisation fraction in the direction of these pulsars.

This paper is structured as follows: Section~\ref{sec:data} gives details of the \hi\ absorption and 3D dust maps used in the study; Section~\ref{sec:methods} summarises calculations of derived quantities and how absorbing CNM is isolated from the larger 3D dust maps; Section~\ref{sec:results} describes the general properties of the derived 1D density profiles towards each pulsar and how we associate individual density components to CNM components in the \hi\ absorption spectra, compares 2D column density images from the dust maps and \hi\ emission (tracing the diffuse WNM), and places constraints on the thermal pressure and ionisation fraction for each LOS; Section~\ref{sec:discussion} places these constraints in context with  predictions from numerical simulations 
and independent observational estimates demonstrates how similar measurements can inform observations of TSAS; and Section~\ref{sec:conclusions} reiterates our conclusions and discusses future applications and improvements of our methods. 

\section{Data}\label{sec:data}

\subsection{3D Dust Maps}\label{subsec:ntot_from_3d_dust}

We use the three-dimensional dust map of \citet{edenhofer24} that are sampled at 516 distance bins spanning 69 pc to 1.25 kpc radially from the position of the Sun. The distance resolution ranges from 0.4 pc at a distance of 69 pc to 7 pc at 1.25 kpc. These maps were constructed by leveraging the stellar distances from parallax measurements with BP/RP spectral data from \textit{Gaia} DR3 \citep{zhang2023} and the dust optical extinction towards approximately 220 million sample stars. Due to artificially high dust extinction values from the positivity prior from \cite{zhang2023}, the inner 69 pc were removed from the maps provided by \citet{edenhofer24}. The posterior inference from \citet{edenhofer24} results in 12 realisations of the dust extinction discretised to HEALPix spheres at logarithmically spaced distances. We utilise these 12 realisations to estimate our uncertainties in the derived line-of-sight (LOS) properties (see Section~\ref{subsec:los_quantities}). We emphasise here that estimated total neutral hydrogen number densities ($n_{\rm tot}$) from the 3D dust maps are strict lower-limits since our interpolation to a Cartesian $xyz$-grid (see Sections~\ref{subsec:los_quantities} and~\ref{subsec:column_density_images}) effectively average the smaller-scale fluctuations over 1 pc$^{3}$ voxels when interpolating the discretised HEALPix spheres to Cartesian grids.

\begin{figure*}
\centering
 \includegraphics[width=\linewidth]{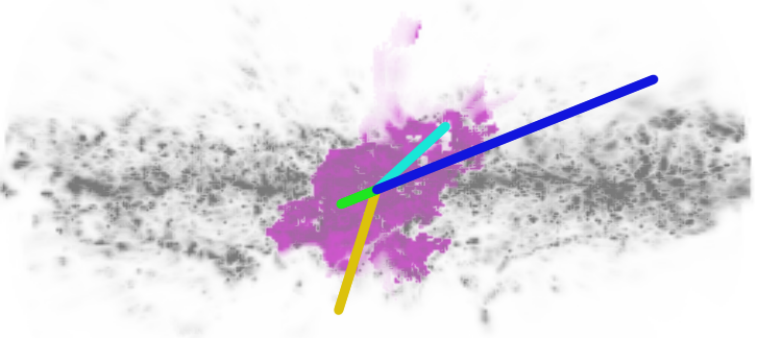}
\caption{The lines of sight from the Sun to each pulsar through the 3D dust distribution from \citet{edenhofer24} in grey scale. B0823+26 is shown in yellow, B1133+16 in green, B1929+10 in turquoise, and B2016+28 in dark blue. The Local Bubble model from \citep{oneill2024} is shown in magenta.}
\label{fig:allLOS}
\end{figure*}

\subsection{\hi\ Emission/Absorption Spectra \& Pulsar Properties}\label{subsec:abs_puls_properties}
The 3D dust maps are used in conjunction with multi-epoch \hi\ absorption measurements made with the Arecibo radio telescope by \citet{stanimirovic2010} and \hi\ emission data from the The Galactic Arecibo L-band Feed Array HI (GALFA-HI) Survey \citep{peek2011, peek2018} measured towards several pulsars---see \citet{stanimirovic2010} for specific details on how these spectra were constructed. The \hi\ absorption spectra were decomposed into individual cloud components employing the Gaussian decomposition method from \citet{heiles2003} to derive $T_{\rm s}$ for each component, the CNM \hi\ column density ($N_{\hi, \rm CNM}$), WNM \hi\ column density ($N_{\hi, \rm WNM}$), and the total \hi\ column density ($N_{\hi, \rm tot}=N_{\hi, \rm CNM}+N_{\hi, \rm WNM}$) along each LOS. 

We study the sightlines towards four pulsars that are within the volume of the dust maps: B0823+26, B1133+16, B1929+10, and B2016+28. \citet{stanimirovic2010} showed that \hi\ absorption and emission spectra towards these pulsars have multiple Gaussian components, indicating that there are multiple CNM clouds along the line of sight moving with different velocities: B1133+16 has two, B1929+10 has three, and B2016+28 has five CNM components. B0823+26 was determined to have only a single component in $\hi$ absorption. In the direction of B1929+10, \citet{stanimirovic2010} noted that the deepest velocity component centred at a velocity of 4.8 \kms\ showed optical depth variability across four epochs, indicating the presence of TSAS on the scale of $\sim$10s of AU when accounting for the proper motion of this pulsar in this specific LOS. 

The pulsar properties are obtained from the ATNF pulsar database \citep{manchester2005} and summarised along with the derived properties for each \hi\ absorption/emission components in Table~\ref{tab:pulsar_properties}.  

\subsection{Model of the Local Bubble}\label{subsec:lb_model_data}
\citet{oneill2024} constructed a model of the Local Bubble by applying a custom peak-finding algorithm to identify inner and outer edges within the larger 3D dust volume of \citet{edenhofer24}. They found the Local Bubble is irregularly shaped and highly porous, indicating it is supernovae-driven superbubble structure that has burst to create a local chimney that funnels ISM material to the Galactic Halo. They found that several well-known molecular clouds and dust features lie on the surface of the Local Bubble. We use the mean shell differential extinction interpolated onto a heliocentric Cartesian grid\footnote{available at \href{https://dataverse.harvard.edu/dataset.xhtml?persistentId=doi:10.7910/DVN/INB1RB}{10.7910/DVN/INB1RB}} to investigate whether the CNM traced by the \hi\ absorption detections from \citet{stanimirovic2010} towards our four pulsars is associated with any structures on the wall of the Local Bubble.

\section{Methods}\label{sec:methods}

\subsection{Derived Line-of-Sight Quantities}\label{subsec:los_quantities}

We use the {\tt Edenhofer2023Query} tool from the publicly available {\tt dustmaps} Python package to extract one-dimensional profiles of the dust extinction from each posterior realisation towards four pulsars with associated \hi\ absorption spectra. 

We set the distance resolution for each extracted profile to be interpolated to 1 pc. Each dust extinction sample along the LOS provides a measurement of the differential extinction ($A'_{\rm ZGR23}=dA_{\rm ZGR23}$/1 pc), where ZGR23 denotes that these extinction estimates are leveraged from the distances and extinctions in the catalogue provided by \citet{zhang2023}. \citet{oneill2024} assumed a constant conversion factor between $A'_{\rm ZGR23}$ and total neutral hydrogen volume density. Following \citet{oneill2024}, we convert the differential extinction into units of total neutral hydrogen volume density by noting that, on parsec scales, integrated extinction can be calculated as the sum of unsmoothed differential extinction,
\begin{equation} A_{\rm ZGR23} =\sum_{i} A'_{\rm {ZGR23},i}dr_i 
\end{equation}
where $dr_{i}$ = 1 pc for all distance bins. We convert to Gaia G-band extinction $A_{\rm G}$ at $\lambda$ = 673 nm \citep{jordi2010}) using the extinction curve from \cite{zhang2023}: 
\begin{equation} 
A_G = 2.0407A_{ZGR23}.
\end{equation}
Then, we assume a constant ratio of extinction to total neutral hydrogen column density, $A_{\rm G}/N_{\rm H}$ = 4$\times10^{-22}$ mag cm$^{2}$ \citep{draine2003, draine2009}, where $A_{\rm G}$ is the extinction of the $G$ band for \textrm{Gaia} and $N_{\rm tot}$ is total neutral hydrogen column density (atomic + molecular contribution); that is, $N_{\rm tot}=N_{\rm HI}+2N_{\rm{H}_2}$). The total neutral hydrogen volume density ($n_{\rm tot}$) at each distance bin along a given LOS can then be derived from extinction:
\begin{equation}
    \begin{split}
     n_{i} & = \frac{1}{A_{G}/N_{H}} \left(\frac{dA_{\rm {G},i}}{1 pc}\right) \left(\frac{da_{i}dr_{i}}{dv_{i}}\right) \\
     & =  \left(\frac{1}{4\times10^{-22}}\right) \left(\frac{2.0407}{3.086\times10^{18}}\right) \rm{A'_{ZGR23,i}}\\
     & = 1653~\rm{cm}^{-3}  A'_{ZGR23,i},
     \end{split}
\end{equation}
where $da_i$ is the projected physical area of the pixel in distance bin $i$, $dr_i$ is the radial separation between distance bin $i$ and $i+1$, and $dv_i = da_idr_i$ is the volume spanned between distance bin $i$ and $i+1$. The factor of $3.086\times10^{18}$ accounts for the conversion between cm and parsecs. \citet{oneill2024} derived a scaling factor equal to 1653 pc mag$^{-1}$ cm$^{-3}$. However, direct measurements of the neutral gas-to-dust ratio $N_{\rm HI}/A_V$ provided by \citet{ONeil2026} suggest a factor of almost two-dex variation in this quantity in a sample of 519 morphologically-matched clouds, indicating that a constant conversion factor that is not a robust assumption. As our analysis requires a scaling factor to estimate the total neutral hydrogen volume density, we adopt the suggested conversion factor of 3002 pc mag$^{-1}$ cm$^{-3}$ from \citet{ONeil2026} to convert the \cite{edenhofer24} 3D dust map to $n_{\rm tot}$. We emphasise that this relies on a \textit{strong} assumption for a fixed ratio between the extinction and total neutral hydrogen column density. Furthermore, as this value pertains only to the atomic hydrogen, this is also likely an underestimate for considering the total amount of hydrogen. The uncertainties for the subsequent volume densities presented in this work only include statistical uncertainties from analysing the 12 available realisations of the dust maps and neglect systematic errors associated with a fixed extinction-to-volume density ratio.

\begin{figure*}
    \centering
    \includegraphics[width=\columnwidth]{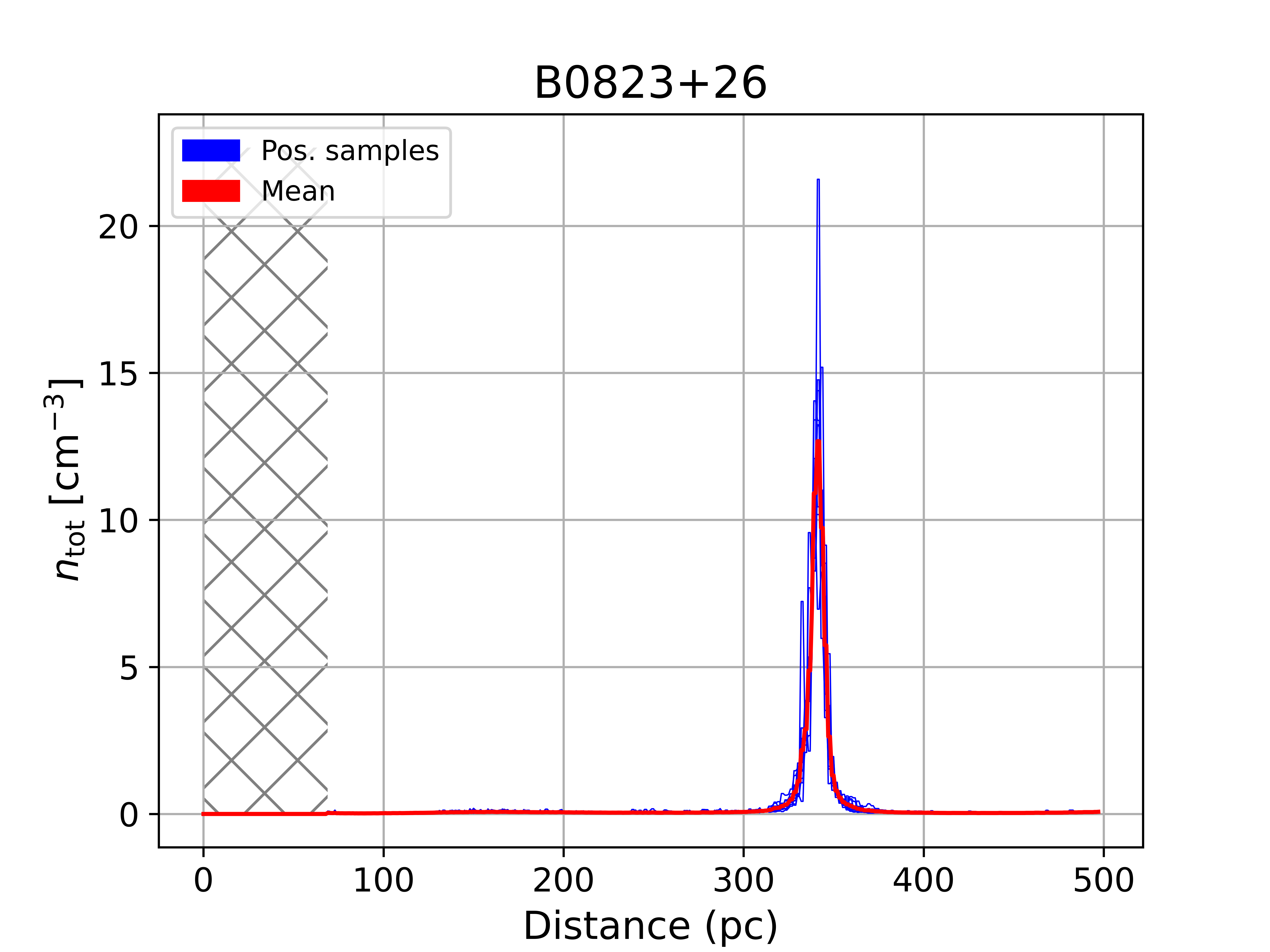}
    \includegraphics[width=\columnwidth]{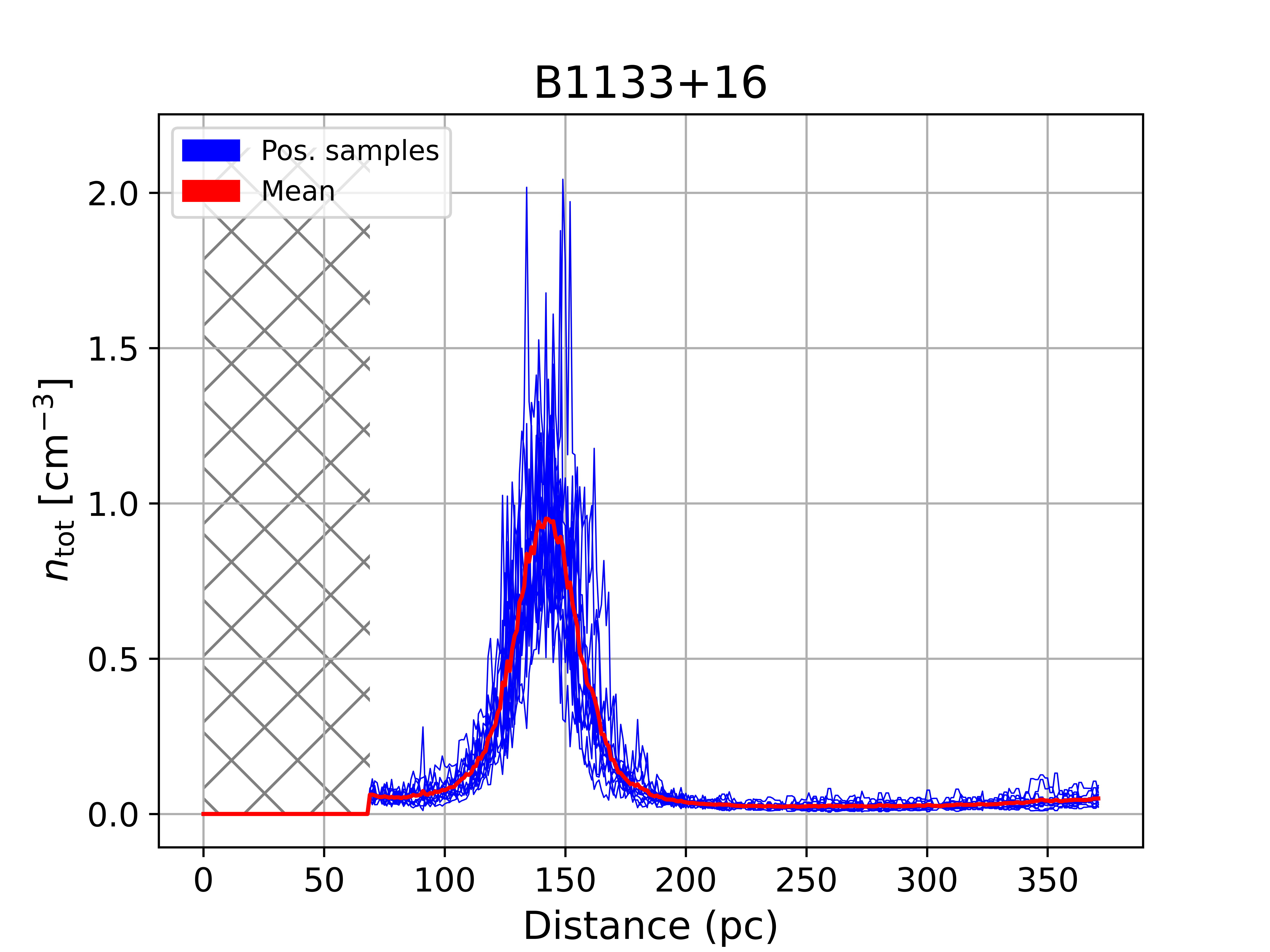}
    \includegraphics[width=\columnwidth]{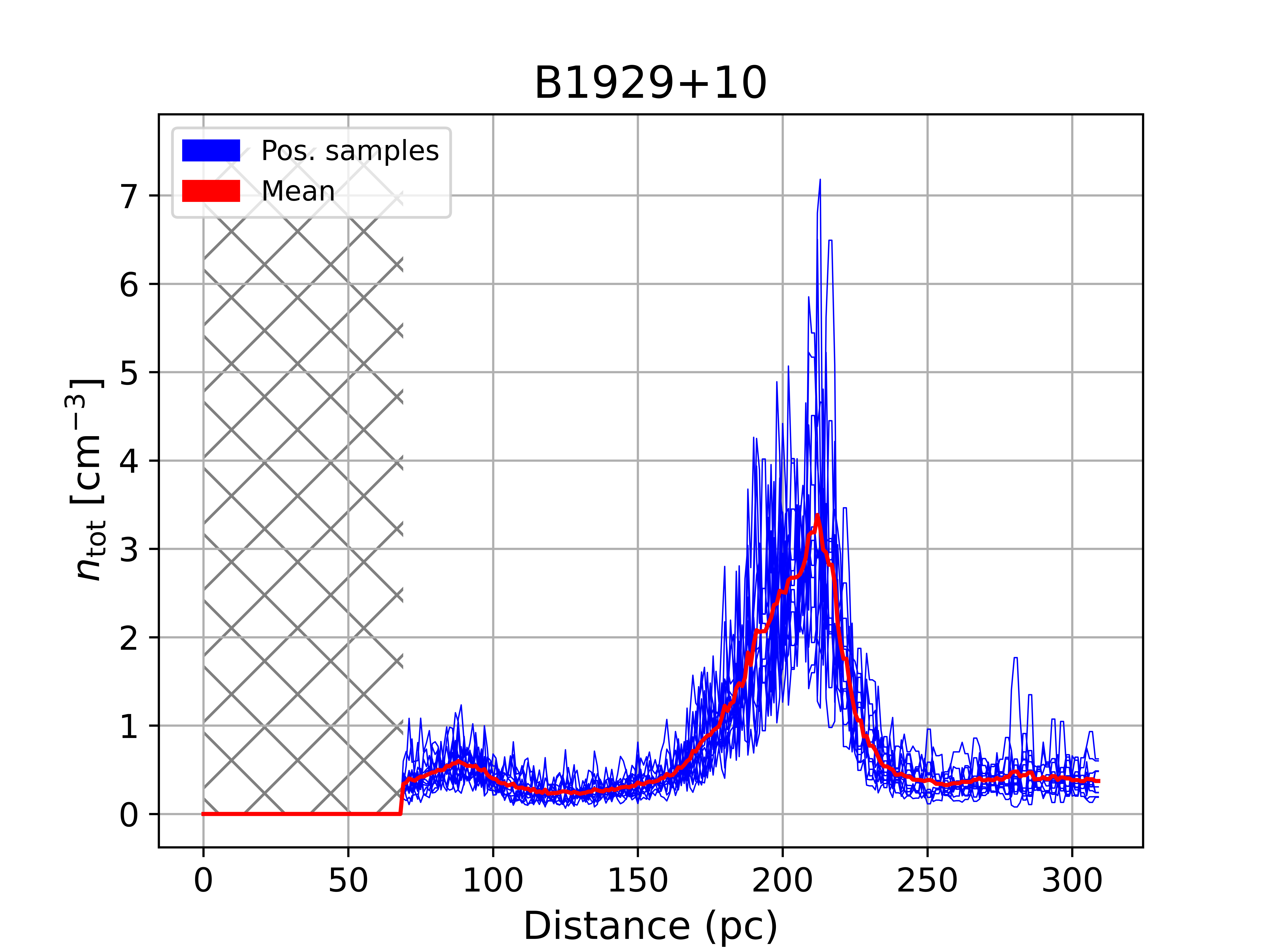}
    \includegraphics[width=\columnwidth]{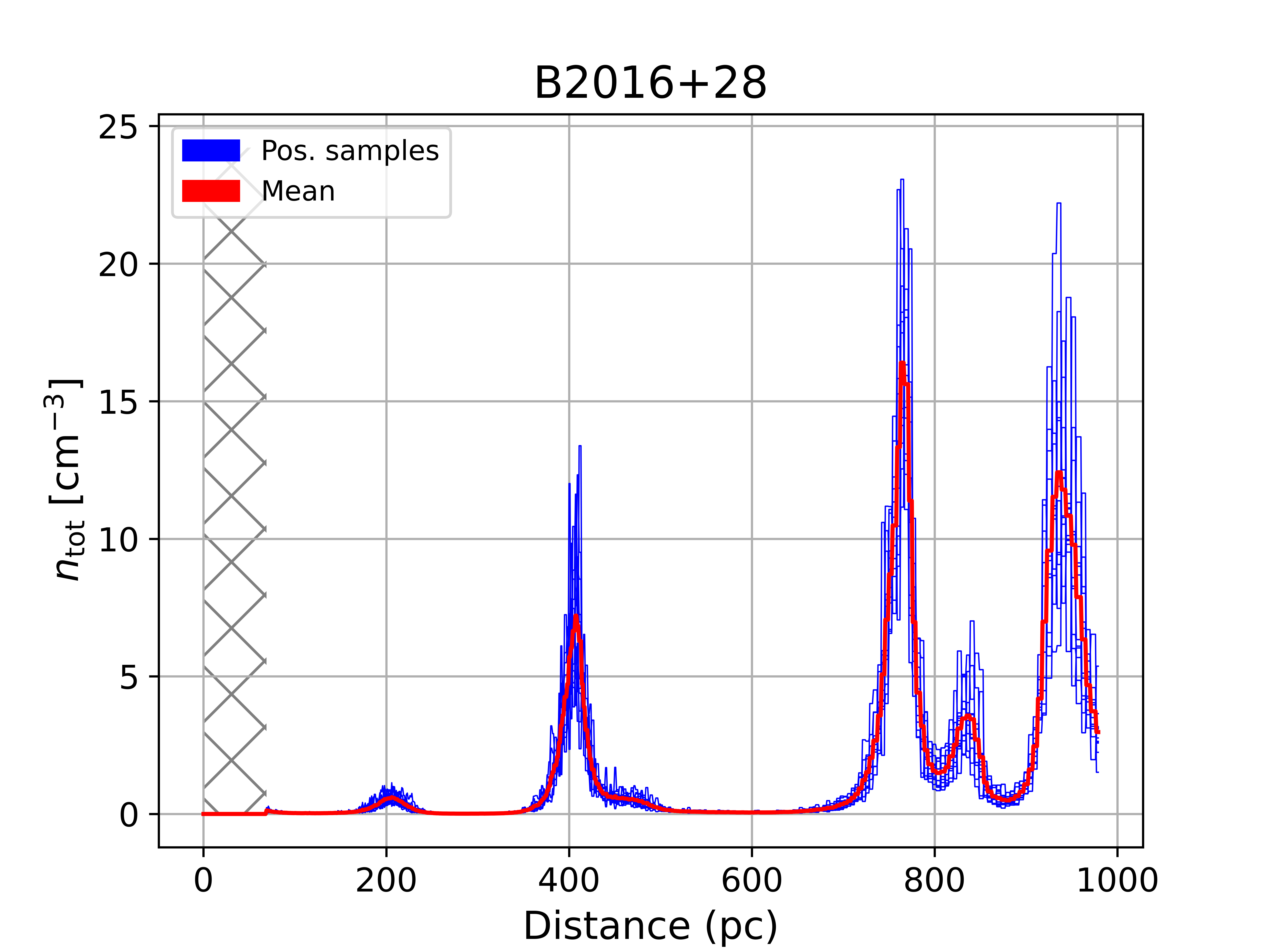}
    \caption{Profiles of $n_{\rm tot}$ towards each pulsar as a function of distance. The profiles extracted from each of the 12 realisations of the dust map posterior samples are shown in blue, and the averaged profile computed using Equation~\ref{eq:ntot_mean} are shown in red. The hatched shaded region denotes the inner 69 pc where there are no data in the dust map posterior samples.}
    \label{fig:comprehensive_density_profiles}
\end{figure*}

The lines of sight from the Sun to each pulsar through the 3D dust distribution are shown in Figure 1.
We estimate the scatter in several derived LOS quantities by extracting a $n_{\rm tot}$ profile from each of the 12 available realisations of the \citet{edenhofer24} dust maps. Figure~\ref{fig:comprehensive_density_profiles} shows the individual (each realisation) and mean $n_{\rm tot}$ profiles towards each pulsar. Each sightline shows notable peaks in $n_{\rm tot}$, indicating the presence of discrete structures along the LOS. For example, the density profiles towards B0823+26 intercept a structure with a maximum $n_{tot}$ at $\sim$330 pc. We also note the position of where the maximum density along the LOS is reached found using a simple argument maximum call to the profile, along with a measure of the Full Width at Half-Maximum (FWHM) identified by eye for each component; the uncertainties are the standard deviations in these measurements across the twelve posterior samples. Our goals are to characterise and isolate the structures along the LOS to relate them to derived properties of the \hi\ absorption. 

We calculate the mean total neutral hydrogen volume density $\langle n_{\rm tot} \rangle$ along each LOS by integrating each of the 12 individual $n_{\rm tot}$ profiles extracted from the posterior samples and dividing by the distance ($d$) using 
\begin{equation}\label{eq:ntot_mean}
\langle n_{\rm tot} \rangle=\frac{1}{d}\int_{0}^{d} n_{\rm tot}\Delta d,
\end{equation}
where $\Delta d$ = 1 pc, and taking an unweighted average. The uncertainty is then the standard deviation across the twelve posterior samples. Similarly, we compute the mean maximum total neutral hydrogen volume density ($\langle n_{\rm tot,max}\rangle$) by averaging the global maximum of the values along each $n_{\rm tot}$ profile and take the uncertainty to be the standard deviation. The mean and associated uncertainty in the total neutral hydrogen column density is computed respectively as the average of
\begin{equation}\label{eq:Ntot_mean}
\langle N_{\rm tot} \rangle=\int_{0}^{d} n_{\rm tot}\Delta d
\end{equation}
and standard deviation across all 12 realisations of the $n_{\rm tot}$ profiles. Finally, we calculate the mean electron density along each LOS towards our pulsars by dividing the $DM$ values in Table~\ref{tab:pulsar_properties}, propagating the statistical uncertainty. These LOS quantities are summarised in Table~\ref{tab:los_properties}. We note that the density values should be considered lower limits as they are interpolated to 1 pc$^3$ voxels.

\subsection{Column Density Images}\label{subsec:column_density_images}
We aim to identify possible absorbing gaseous structures to place limits on the physical properties of the ISM along each LOS. We use the {\tt interp2lbd.py} script provided by \citet{edenhofer24} to interpolate the mean posterior 3D dust map---scaled to units of $n_{\rm tot}$---to a sub-volume spanning $10^{\circ}\times10^{\circ}$ in Galactic Longitude ($l$) and Galactic Latitude ($b$) centred on the location of each pulsar and out to the distance of each pulsar in 1 pc bins (see Table~\ref{tab:pulsar_properties}). We generate the two-dimensional total neutral hydrogen column density image by integrating the $n_{\rm tot}$ values along the entire LOS to each pulsar.

In latter sections, we make a comparison with \hi\ emission from an equivalent field of view from publicly available GALFA-\hi\ data cubes\footnote{\url{http://purcell.ssl.berkeley.edu}}. We reproject the \hi\ emission cubes from J2000 equatorial coordinates to Galactic coordinates using the {\tt reproject} package available in {\tt astropy} and compute the \hi\ column density under the optically-thin assumption
\begin{equation}\label{eq:hi_column_density}
N_{\rm HI}=1.82\times10^{18}\int^{v_{\rm max}}_{v_{\rm min}}~T_{\rm b}~dv ~\textrm{cm}^{-2},
\end{equation}
where $dv$=0.74 \kms\ and $T_{\rm b}$ is the brightness temperature along each line of sight, and $v_{\rm max}$ and $v_{\rm max}$ and the maximum and minimum velocity range over which we integrate. 

\subsection{Isolating Absorbing Structures}\label{subsec:isolation_methods}
\begin{figure*}
\centering
    \includegraphics[scale=0.4, width=\columnwidth]{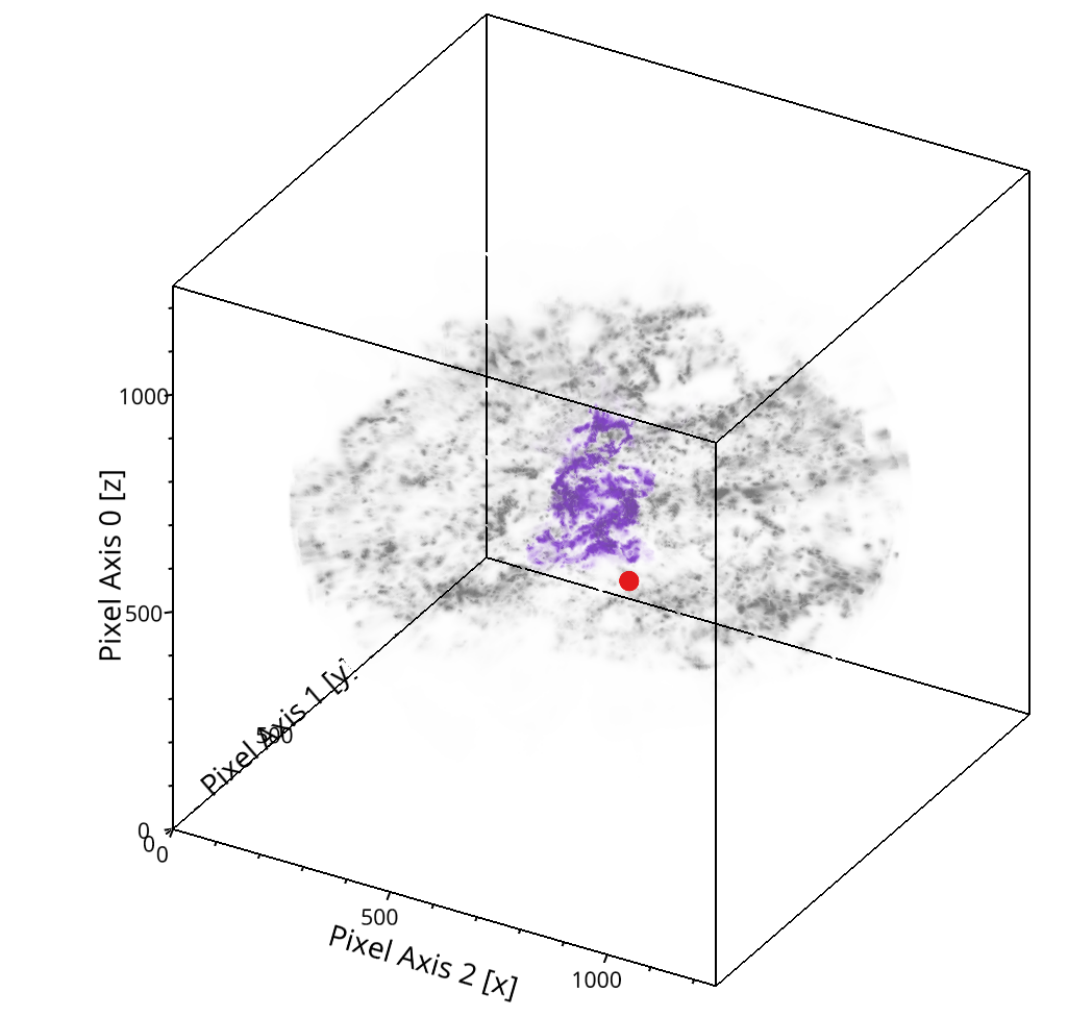}
    \includegraphics[width=\columnwidth]{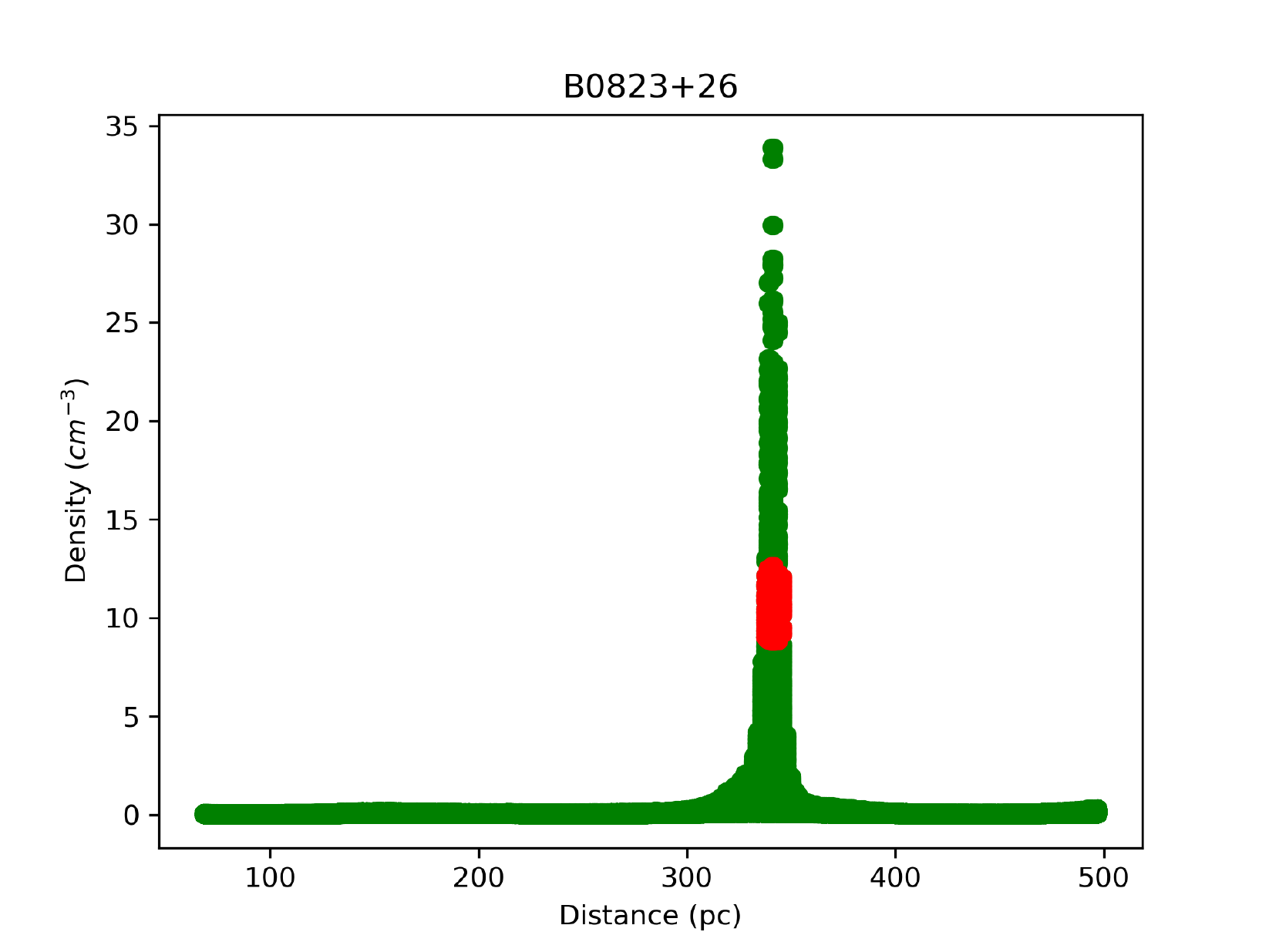}
    \caption{Summary of our isolation of 3D dust structure associated with the main density peak for B0823+26. The left image shows the 3D isolated structure in red, the LB model is highlighted in purple while the 3D dust data from \citet{edenhofer24} are shown in gray (visualisation from Glue). The right image shows the range of distances and the hydrogen density values corresponding to the isolated structure.}
    \label{fig:isolation_summary}
\end{figure*}

We aim to characterise the three-dimensional structure associated with the CNM responsible for the observed \hi\ absorption. Constraining the mean $n_{\rm tot}$ of the absorbing structure enables estimates of key ISM properties, such as the thermal pressure and ionisation fraction. We also seek to determine whether the absorbing gas resides within, or on the wall of, the Local Bubble (LB), thereby providing a direct comparison with nearby cold clouds such as the Leo Cold Cloud.

To isolate dense structures associated with the absorption, we identify peaks in the reconstructed density profiles and corresponding dusty features in the column density maps along each line of sight using the subset selection tools available within the visualisation software {\tt Glue} \citep{robitaille2017}. We first generate catalogues for the sub-volumes extracted in Section~\ref{subsec:column_density_images} from the mean dust map containing the $(x, y, z)$ Cartesian coordinates, Galactic coordinates, distance (pc), and $n_{\rm tot}$ value of every voxel. These catalogues are then loaded into {\tt Glue}, where we construct a two-dimensional scatter plot of $n_{\rm tot}$ as a function of distance. We define a rectangular subset spanning the FWHM of the density peak (shown by the red profiles in Figure~\ref{fig:comprehensive_density_profiles}) together with the corresponding range of $n_{\rm tot}$ values for the primary absorbing component along the line of sight.

Figure~\ref{fig:isolation_summary} shows these selected voxels mapped back onto the three-dimensional dust reconstruction of \citet{edenhofer24}, allowing us to visualise the location of the absorbing CNM relative to both the Earth and the Local Bubble along the sightline toward B0823+26. In this case, the isolated dense structure lies beyond the Local Bubble, indicating that the CNM detected in \hi\ absorption is not associated with the LB. The right panel shows the scatter plot used to isolate the voxels associated with the absorbing structure. Higher density values that are not highlighted emanate from voxels within the extracted sub-volume but not along the line of sight towards the pulsar.

We emphasise that this procedure is intended as a qualitative visualisation of the absorbing structures rather than a quantitative decomposition. Nevertheless, it provides valuable insight into their three-dimensional locations and morphologies. In the following sections, we use the reconstructed $n_{\rm tot}$ profiles and column density distributions to quantitatively assess the geometry of these structures and their relationship to the Local Bubble.

\section{Results}\label{sec:results}

\begin{table*}[htp]
	\begin{center}
	\caption{Summary of the pulsar properties. The integrated \hi\ column density in the direction of each pulsar is from \citet{stanimirovic2010}.}
	\label{tab:pulsar_properties}
	\begin{tabular}{ccccccccc}
\hline\hline
Pulsar & (l, b) & Distance & Dispersion Measure (DM) & $N_{\hi,\rm tot}$ & Reference \\
\
 & [deg,deg] & [pc] & [pc cm$^{-3}$] & [$10^{20}$ cm$^{-2}$] &  \\ \hline
B0823+26  & (196.96, 31.74) & 500 & 19.4763 & 5.0 & \cite{Bilous2016}/\citet{gwinn1986} \\
B1133+16  & (241.90, 69.19) & 370 & 4.841 & 4.2 & \citet{Bilous2016}/\citet{brisken2002}\\
B1929+10 & (47.380, $-$3.88) & 310 & 3.1832 & 37.2 & \citet{Bilous2016}/\citet{brisken2002} \\
B2016+28 & (68.10, $-$3.98) & 980 & 14.1977 & 47.7 & \citet{srb+15}/\citep{brisken2002} \\
\hline
	\end{tabular}
	\end{center}
    \begin{minipage}{\linewidth}
    The pulsar properties are taken from the ATNF pulsar database \citep{manchester2005}. The references in the final column pertain to the DM and distances, respectively. The uncertainties for the DM values are $<$0.01\%. The \hi\ column density is from \citet{stanimirovic2010}.
    \end{minipage}
\end{table*}

\subsection{Total neutral Hydrogen Volume Density Profiles \& Column Density Images}\label{subsec:ntot_volume_density_profiles}

\begin{figure*}
 \centering
  \includegraphics[width=\textwidth]{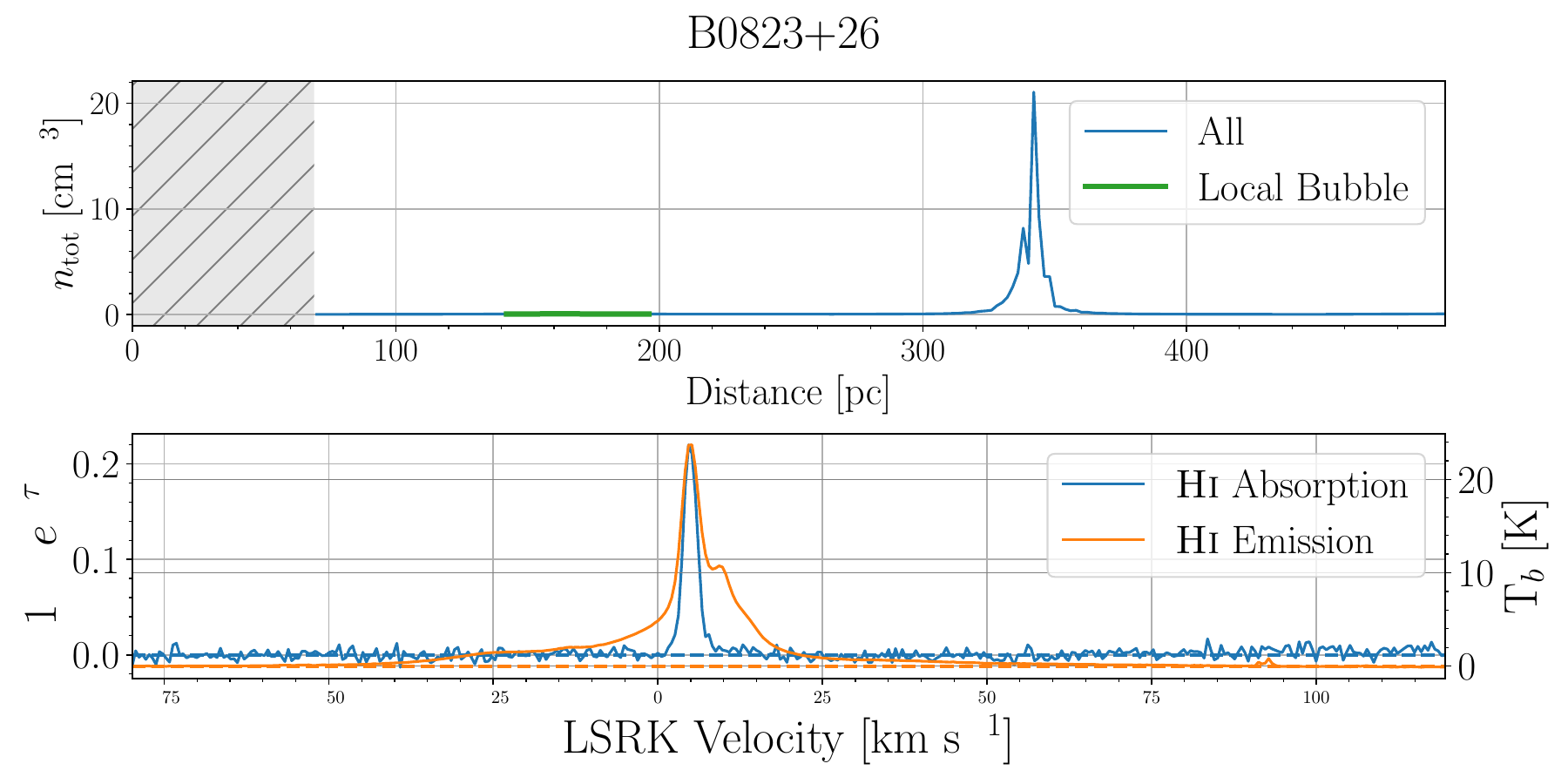}
  \includegraphics[width=\textwidth]{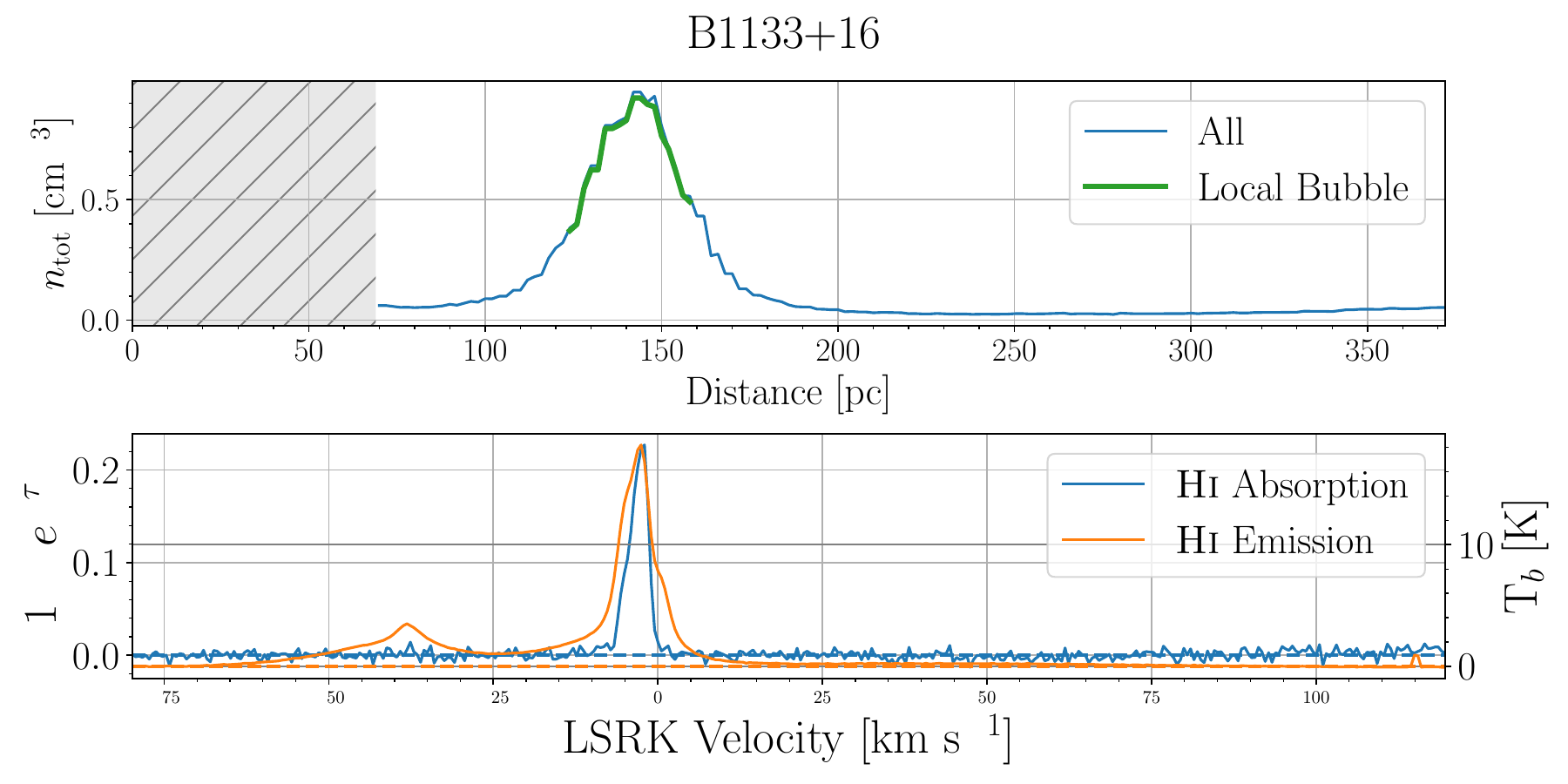}

\caption{Top: The total neutral hydrogen volume density extracted from a single sightline towards each pulsar from the mean cube of the twelve posterior realisations of the 3D dust maps presented in \citet{edenhofer24} as a function of distances. Structure attributed to the Local Bubble according to the model derived by  \citet{oneill2024} is marked in green. The shaded regions represent the inner 69 pc where no data are available. Bottom: The optical depth spectrum (left y-axis) from \hi\ absorption and associated mean \hi\ emission spectrum brightness (right y-axis) for a single sightline towards each pulsar.}
\label{fig:mean_density_profiles}
\end{figure*}

\begin{figure*}
\ContinuedFloat
 \centering
  \includegraphics[width=\textwidth]{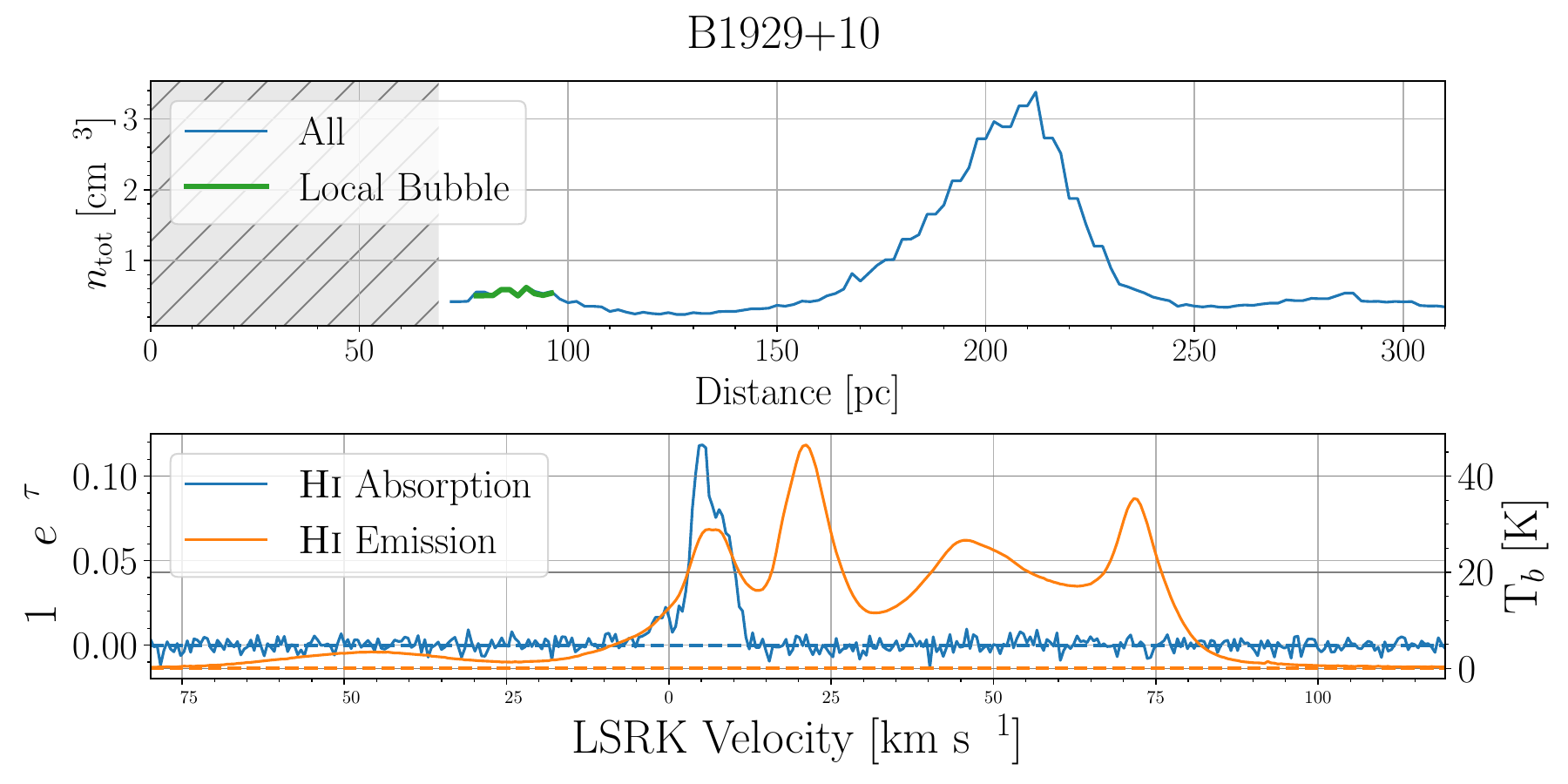}
  \includegraphics[width=\textwidth]{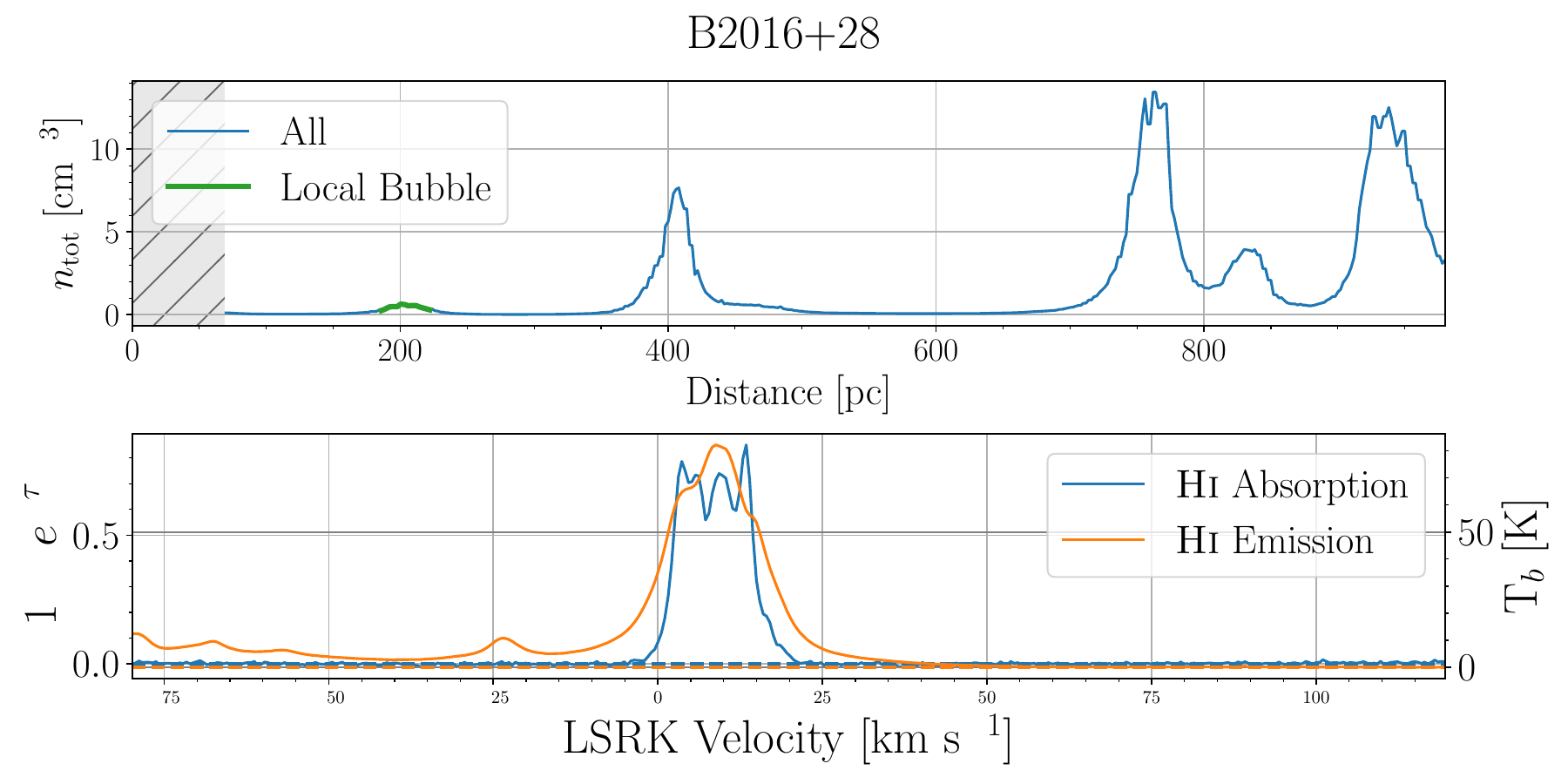}
\caption{\textit{continued}}
\label{fig:mean_density_profiles}
\end{figure*}

Figure~\ref{fig:mean_density_profiles} presents $n_{\rm tot}$ profiles extracted from the mean \citet{edenhofer24} 3D dust cube---the average of the 12 available realisations---toward each pulsar paired with the associated \hi\ emission and absorption along the same sightlines from \citet{stanimirovic2010}. For B1133+16 and B1929+10, these 1D density profiles show relatively weak, broad structures peaking at distances of 140 pc and 210 pc, respectively. In contrast, B0823+26 exhibits a sharp, high-density structure near 350 pc, while B2016+28 displays multiple distinct peaks distributed along the entire LOS. The peak mean volume densities toward B0823+26 and B2016+28 reach approximately 15 cm$^{-3}$---significantly higher than those in the other two directions. These peaks are also more compact, with FWHM values of $\sim$11 pc for B0823+26 and $\sim$20$-$40 pc for B2016+28, compared to broader features (FWHM 30 pc) along the LOS to B1929+10 and B1133+16. Notably, pulsars with multiple density components---B1929+10 and B2016+28---are located closer to the Galactic Plane, suggesting their sightlines primarily sample the warmer ISM concentrated in the Galactic disk.

\begin{figure*}
    \centering
    \includegraphics[width=\columnwidth]{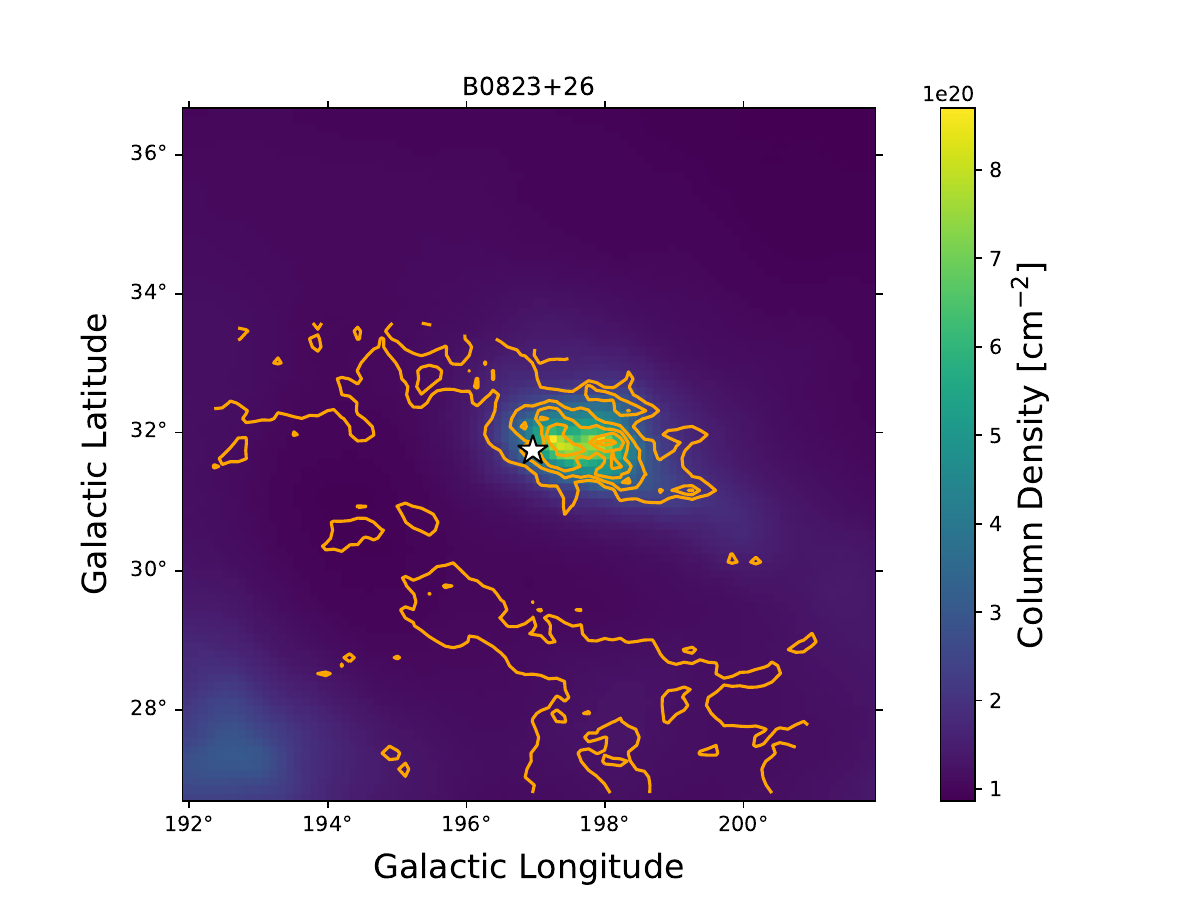}
    \includegraphics[width=\columnwidth]{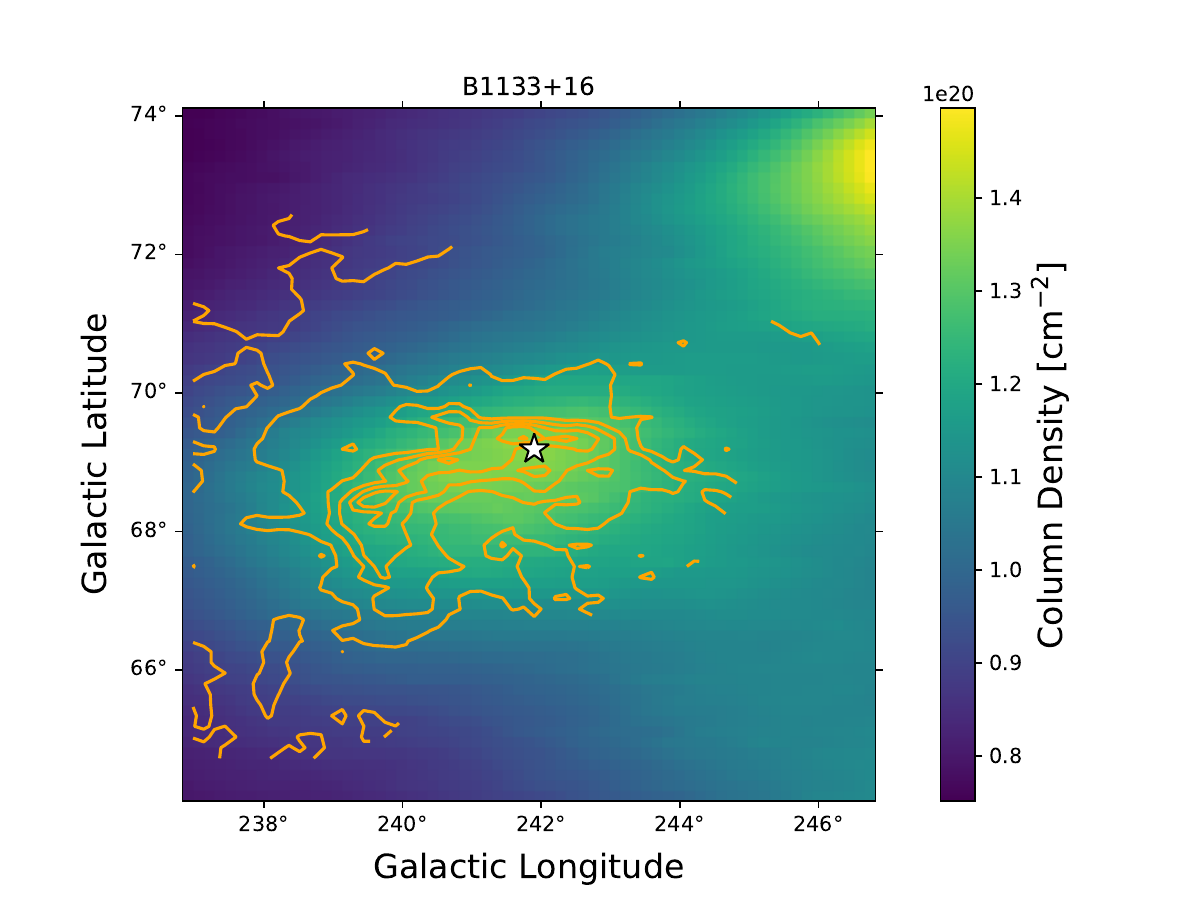}
    \includegraphics[width=\columnwidth]{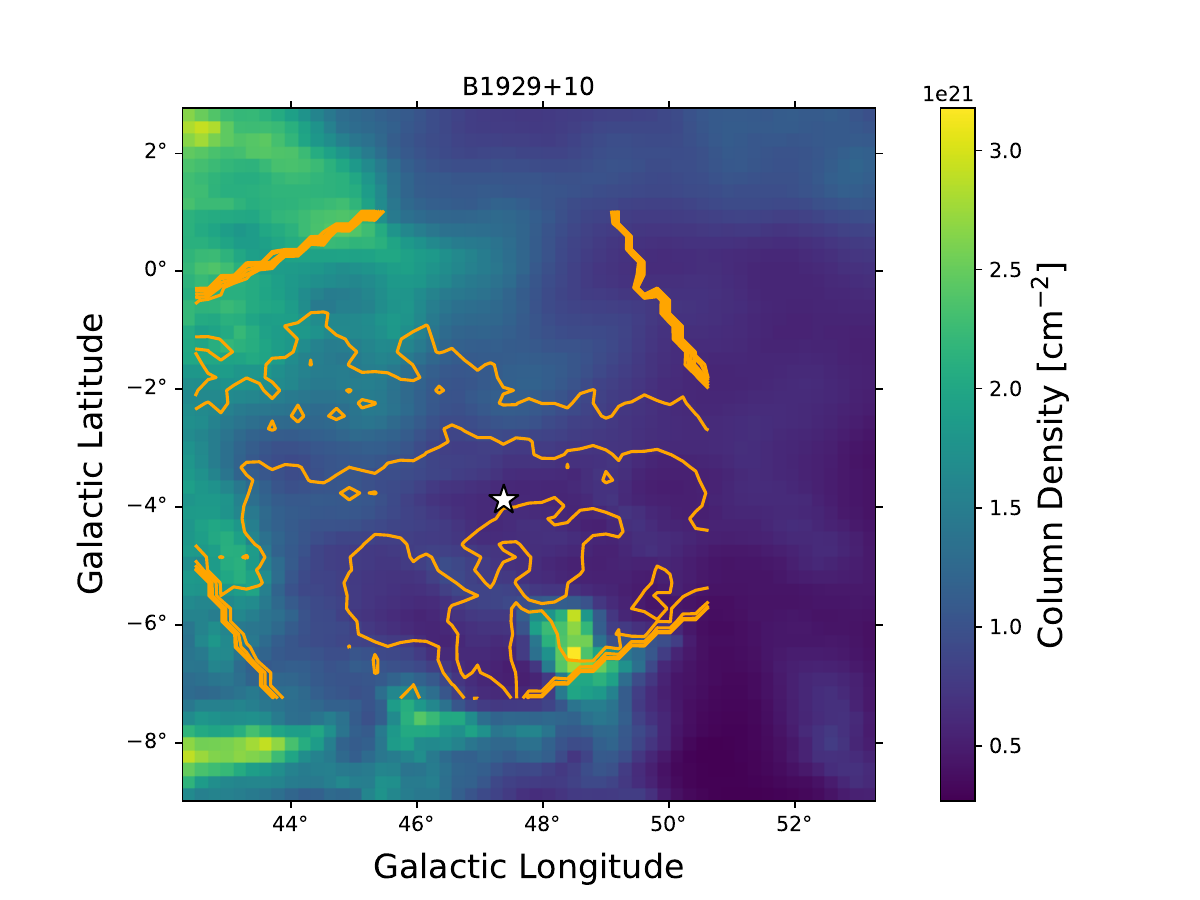}
    \includegraphics[width=\columnwidth]{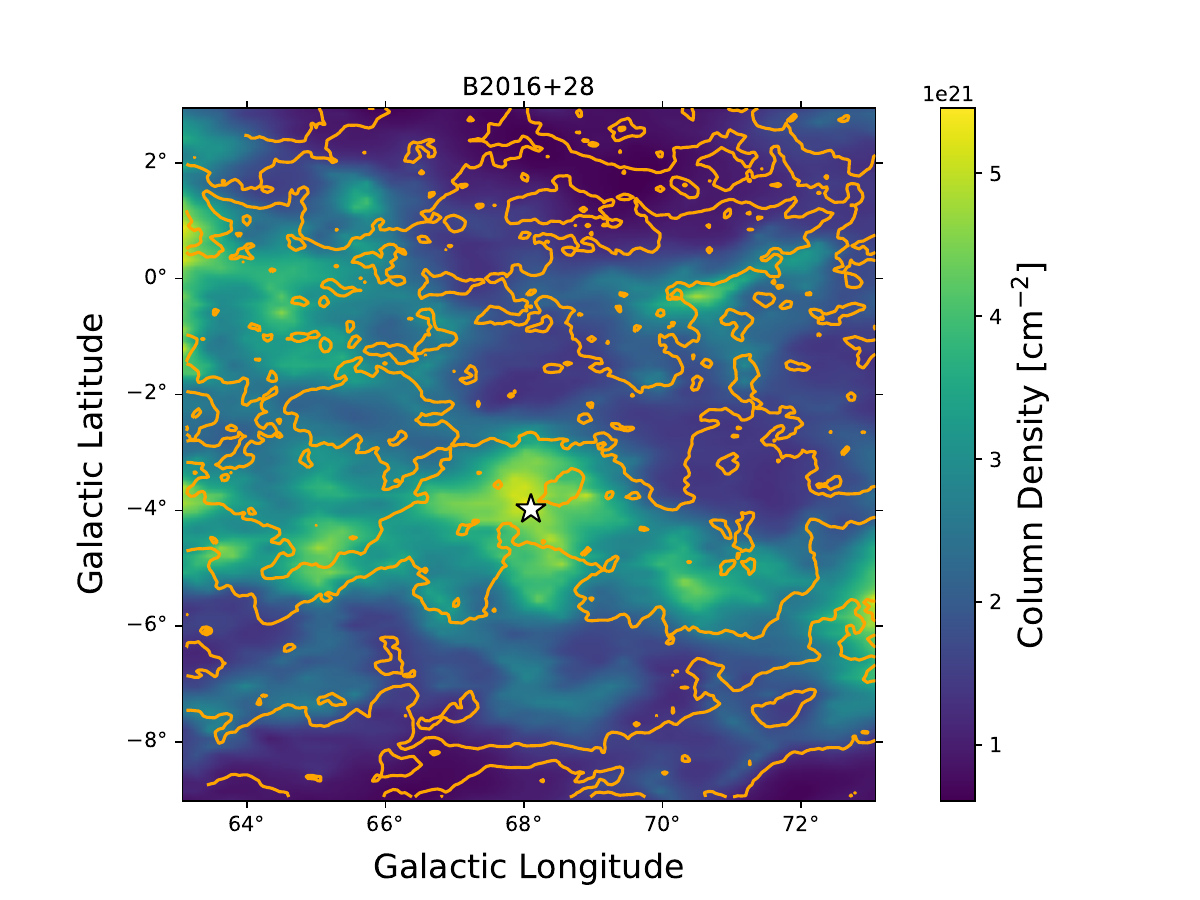}
    \caption{The total neutral hydrogen column density image obtained by integrating our extracted subvolume from the 3D \citet{edenhofer24} dust map along the extent of the line of sight to each pulsar. The orange contours show \hi\ column density computed over the full LSR velocity range of the CNM components identified by \hi\ absorption (see Table~\ref{tab:cnm_derived_properties}) at six equally binned levels relative to the peak value. The location of the pulsar is marked by the white star.}
    \label{fig:N_tot_images}
\end{figure*}

The general characteristics of the 1D density structure aligns with the total neutral hydrogen column density maps in Figure~\ref{fig:N_tot_images} from the extracted subvolumes. The 2D structure along the LOS to the pulsars range from diffuse environments without obvious small-scale structure to clumpy, isolated features. In particular, B1133+16 does not show any small-scale (dust) structure, while the B0823+26 LOS intercepts an isolated structure not associated with a known molecular cloud. In the direction of B1929+10, the total neutral hydrogen column density image shows a filamentary feature several degrees away from the pulsar LOS possessing small-scale, clumpy structure. The structure towards B2016+28 is a combination of underlying diffuse structure extending below the Galactic plane with small-scale structure sprinkled throughout the large-scale feature. 

\citet{stanimirovic2010} suggested that the variability of \hi\ optical depth spectra towards B1929+10 could be associated with cold \hi\ structures on the surface of the wall of the Local Bubble. To connect our observed density structures to the Local Bubble, we extract sightlines from the \citet{oneill2024} model by converting $l$, $b$, and $d$ to $(x, y, z)$ Cartesian coordinates using standard spherical coordinate transformations, which are overplotted in green on the total neutral hydrogen density profiles in Figure~\ref{fig:mean_density_profiles}. We note that typical uncertainties for the distance to LB features from \citet{oneill2024} are $\sim50$ pc , and less than $10$ pc in the dense shell, with a correlation between density and uncertainty (see Figure B2 in \citealt{oneill2024}). Interestingly, the prominent density feature towards B1929+10 does not correspond to any structure associated with the \citet{oneill2024} model of the Local Bubble. In fact, the only prominent total neutral hydrogen structure showing association with the Local Bubble wall is the single peak towards B1133+16, indicating that the majority of the \hi\ absorption comes from structures located more than 100 pc outside of the Local Bubble. Though two weaker density structures in the directions of B1929+10 and B2016+28 at distances of $\sim$70 pc and 200 pc, respectively, are associated with the LB wall.

The line-of-sight towards B0823+26 intersects an interstellar cloud with the highest peak density, $\sim20$ cm$^{-3}$. Previously, \citet{zucker2021} investigated the structure of molecular clouds in the Solar vicinity in three-dimensional dust data \citep{leike2020}. While the LOS extent of the structure in the direction of B0823+26 is in good agreement with the typical depth of nearby molecular clouds ($\sim20$ pc), the peak density of molecular clouds is usually around 30-40 cm$^{-3}$. This suggests that while the structure towards B0823+26 is well defined, it is more diffuse than typical molecular clouds. On the other hand, \citet{zucker2021} note that the three-dimensional dust maps of \citet{leike2020}---which employ a reconstruction methodology similar to that used by the \citet{edenhofer24} maps adopted in this work---tend to underestimate the extinction in the densest regions. This bias arises because the extinction toward background stars becomes sufficiently large that the number of detectable stars behind dense clouds is reduced, limiting their usefulness in the reconstruction. Consequently, the molecular cloud identified here may be more representative of the population discussed by \citet{zucker2021} than implied by the inferred total neutral hydrogen volume density profile.
\begin{table*}[htp]
	\begin{center}
\caption{Derived LOS Properties Towards each Pulsar}
	\label{tab:los_properties}
	\begin{tabular}{lccccc}
\hline\hline
Pulsar & $\langle n_{\rm tot} \rangle $& $\langle n_{\rm tot,max} \rangle$&  $\langle n_e \rangle $&$\langle N_{\rm tot} \rangle$ &$\langle x_e \rangle$\\ 
\
 & [cm$^{-3}$]& [cm$^{-3}$]&  [cm$^{-3}$]&[$10^{20}$ cm$^{-2}$] &\\
\hline
B0823+26    & $0.34\pm0.04$&$13\pm4$& 0.0381 &$4.5\pm0.4$ &$0.11\pm0.01$\\ 
B1133+16    & $0.14\pm0.02$& $1.4\pm 0.5$& 0.0131 & $1.3\pm0.2$ &$0.09\pm0.01$\\ 
B1929+10    & $0.76\pm 0.04$& $3.2\pm 0.6$&  0.0103 &$5.7\pm0.2$ &$0.010\pm0.001$\\
B2016+28    & $1.64\pm0.05$& $18\pm2$&  0.0145 & $47\pm1$ &$0.008\pm0.002$\\ 
\hline
	\end{tabular}
        \begin{minipage}{\linewidth}
    Columns (2)$-$(6) respectively provide mean total neutral hydrogen volume density across the 12 realisations of the dust maps, the mean maximum total neutral hydrogen volume density, the mean electron density derived from the DM and distance values in Table~\ref{tab:pulsar_properties}, the mean total neutral hydrogen column density towards each pulsar across the 12 realisations of the dust maps, and the mean ionisation fraction.
    \end{minipage}
	\end{center}
\end{table*}

\subsection{Correlating \hi\ absorption with $n_{\rm tot}$ profiles}\label{subsec:correlate_abs_with_ntot}

We next seek to correlate the $n_{\rm tot}$ profiles derived from dust observations with the \hi\ absorption in Figure~\ref{fig:mean_density_profiles}. The $n_{\rm tot}$ profiles are qualitatively more similar to \hi\ absorption spectra than the \hi\ emission, which contain contributions from both cold and warm \hi\ structures located in front and beyond the pulsars' positions. It is important to note that \hi\ and dust data sets have different resolutions. The \hi\ absorption spectra come from a pencil-sharp beam in the direction of pulsars, while \hi\ emission spectra have angular resolution of $\sim4'$---corresponding to 0.1 pc$-$0.25 pc at a distance of 100$-$200 pc, and the 3D dust data sets are interpolated to a Cartesian grid with a spatial resolution of 1 pc. We summarise the derived CNM and dust structure properties for individual components along the LOS towards each pulsar in Tables~\ref{tab:cnm_derived_properties} and~\ref{tab:dust_derived_properties}, respectively. 

\begin{figure*}
    \centering
    \includegraphics[width=0.40\textwidth]{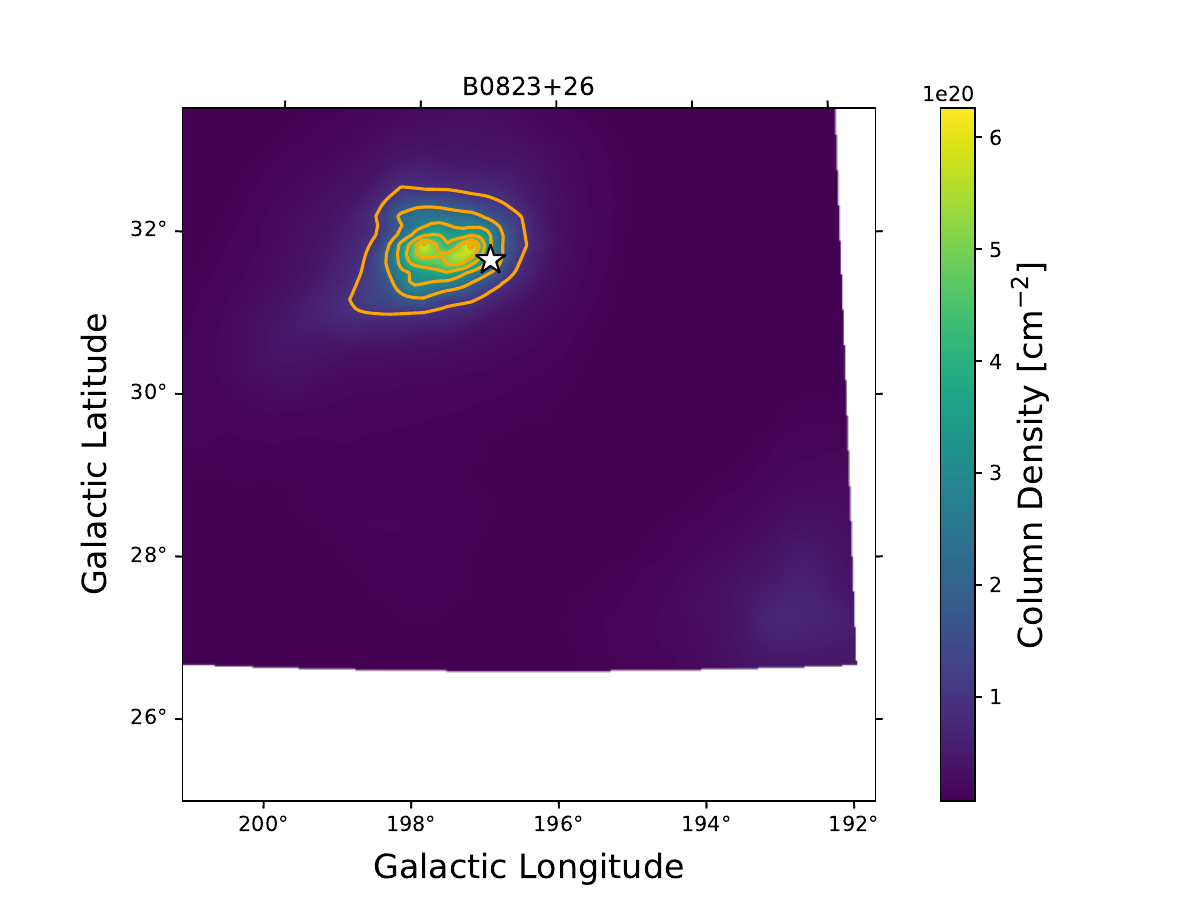}
    \includegraphics[width=0.40\textwidth]{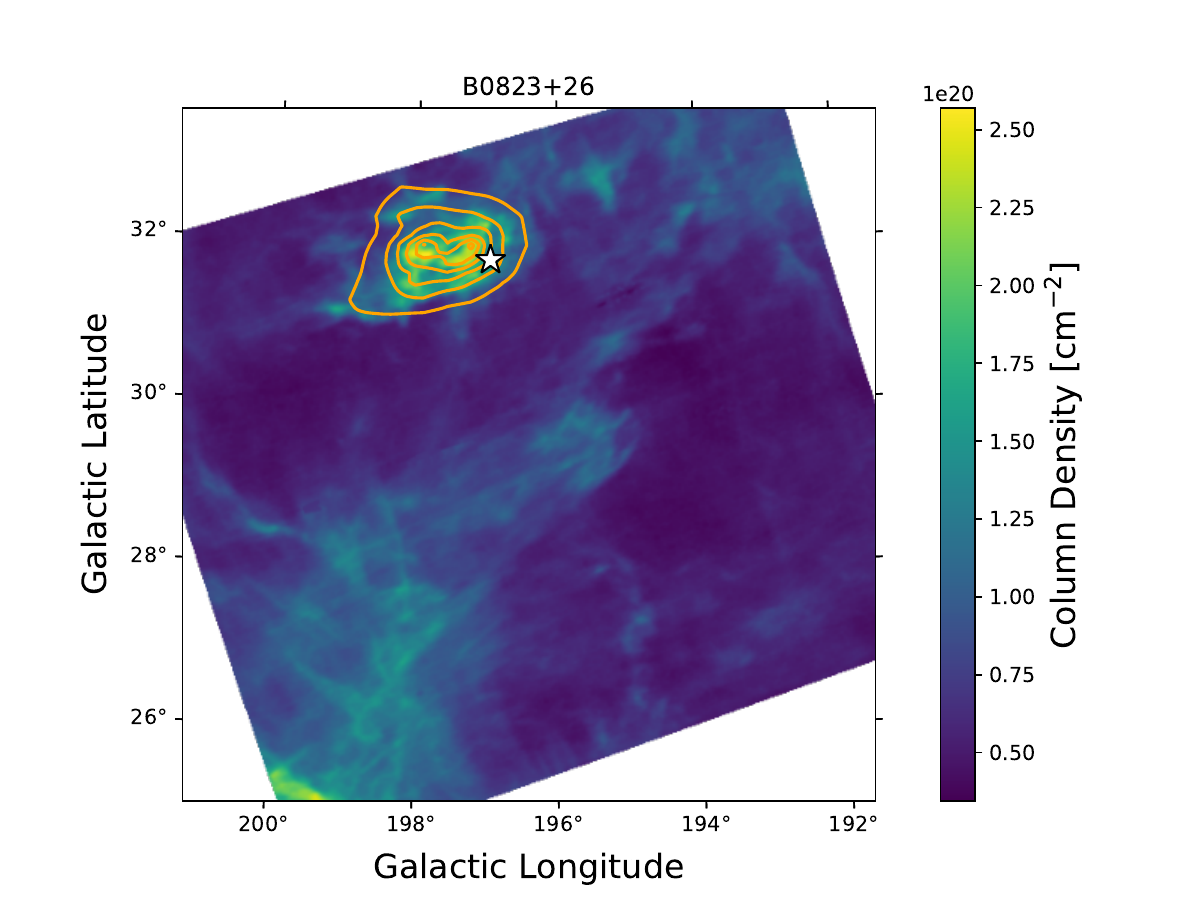}
    \includegraphics[width=0.40\textwidth]{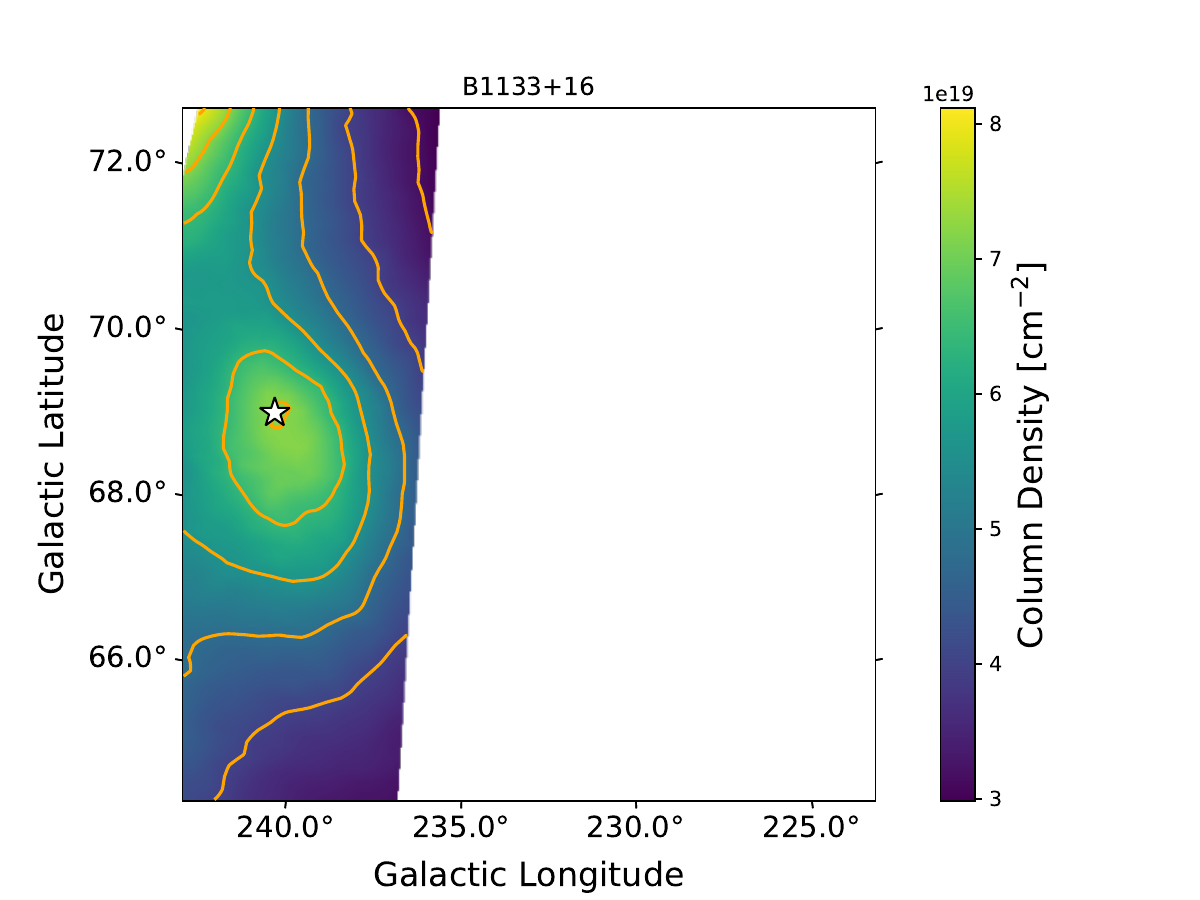}
    \includegraphics[width=0.40\textwidth]{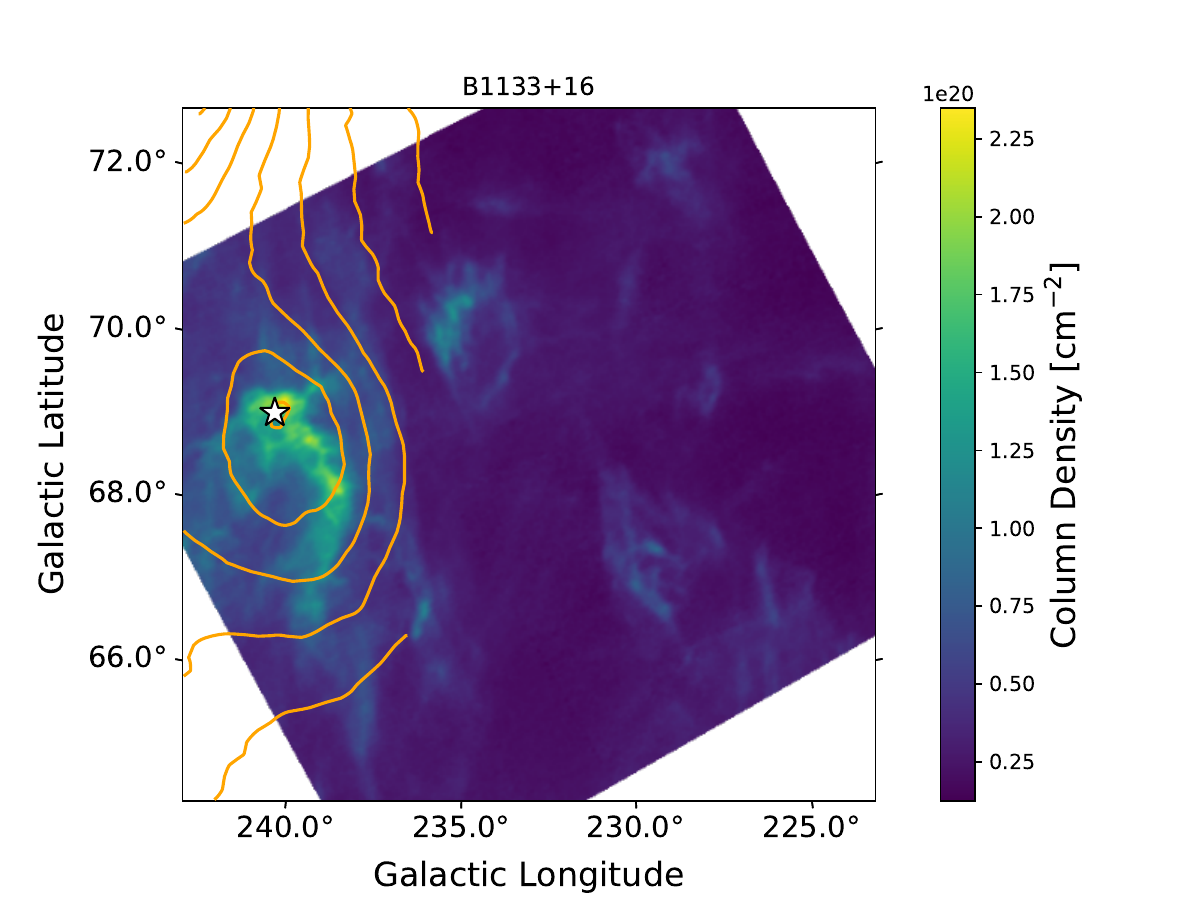}
    \includegraphics[width=0.40\textwidth]{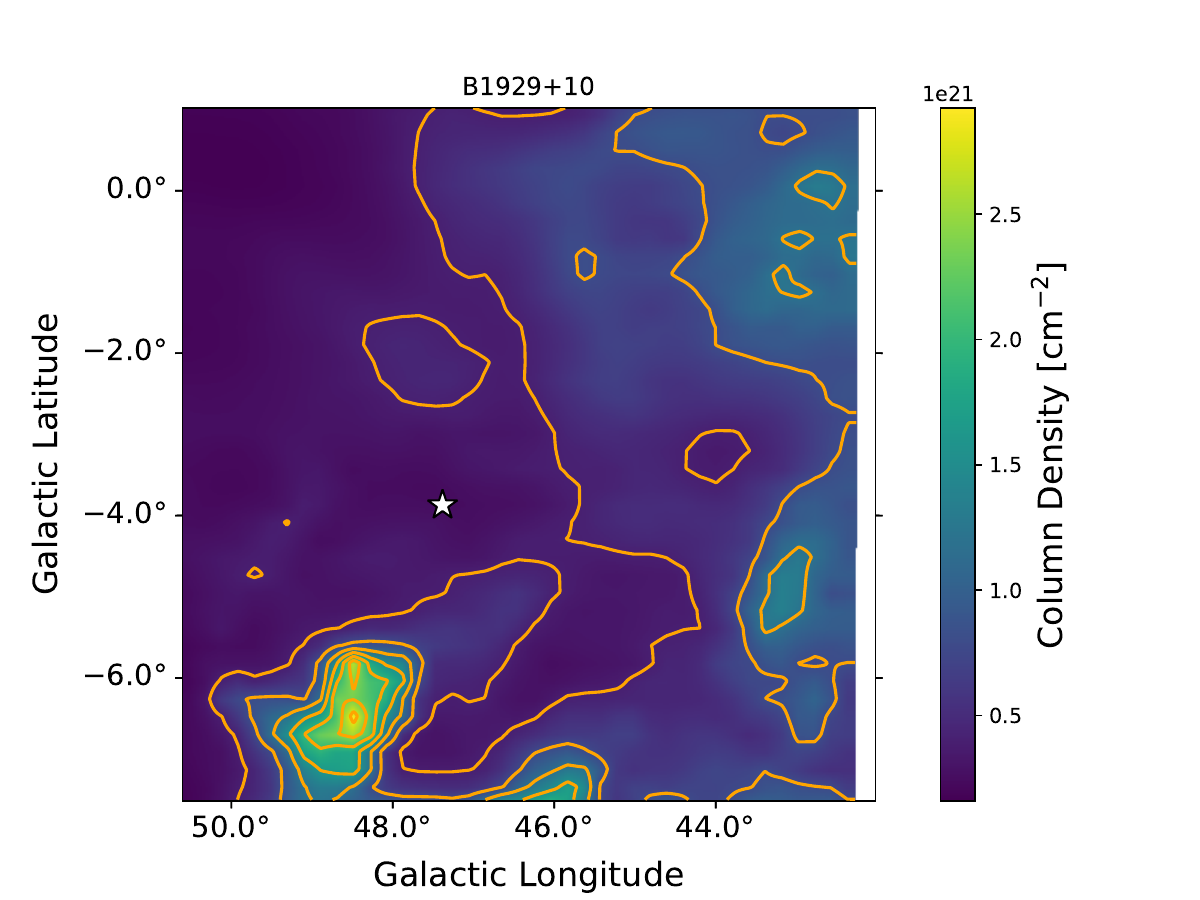}
    \includegraphics[width=0.40\textwidth]{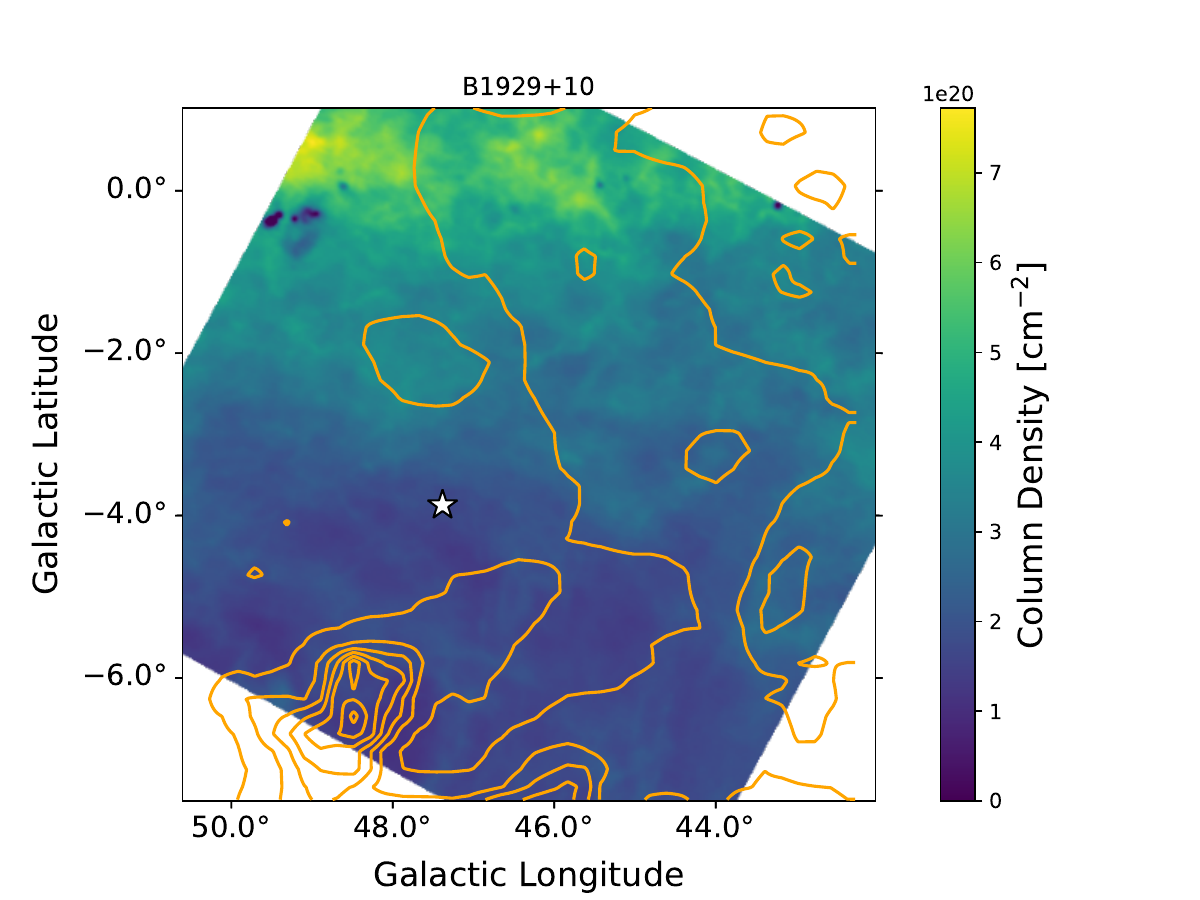}
    \includegraphics[width=0.40\textwidth]{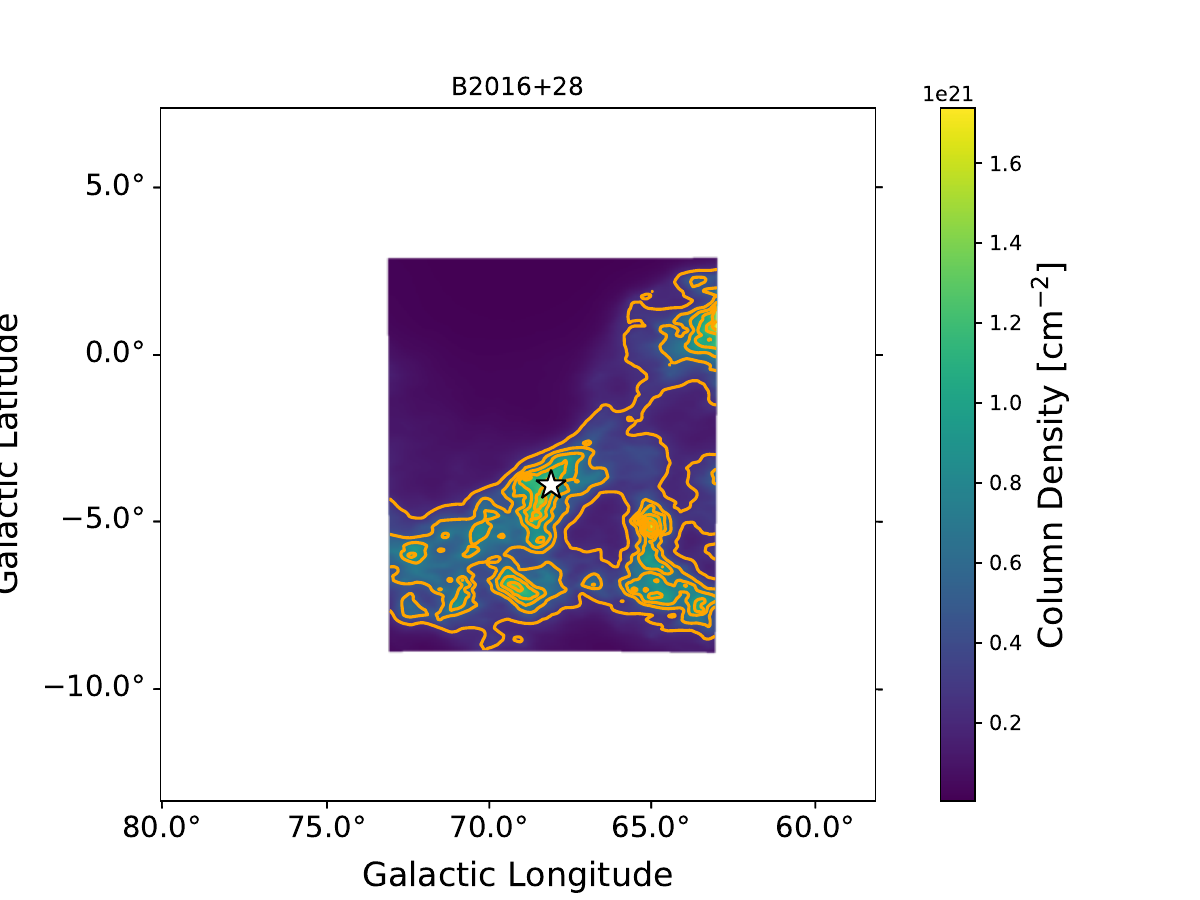}
    \includegraphics[width=0.40\textwidth]{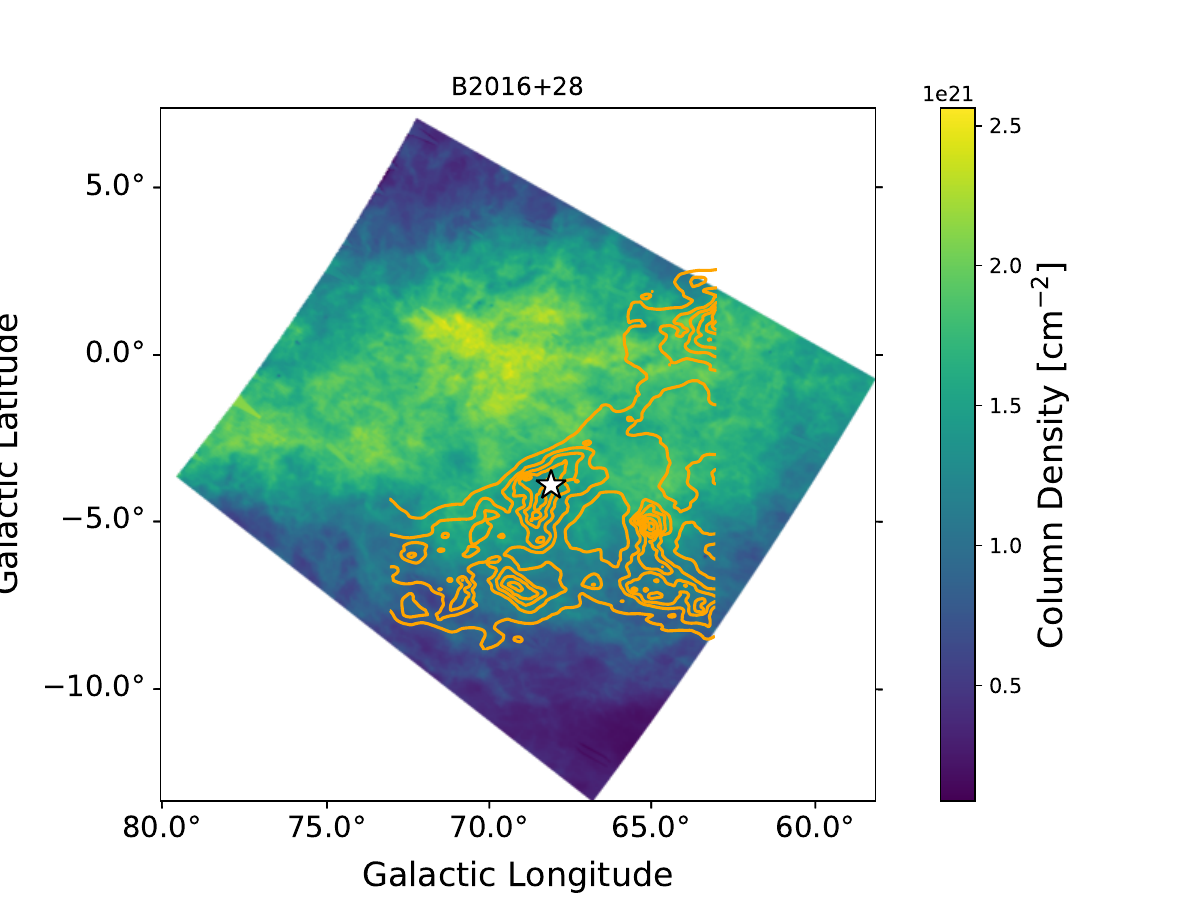}
\caption{Total neutral hydrogen (left column) and \hi\ (right column) column density images towards each pulsar. For each LOS with multiple components (B1929+10 and B2016+28), we show the integration over the FWHM of the strongest density component and integrate over the corresponding velocity component from the \hi\ absorption features (summarised in Table~\ref{tab:dust_derived_properties}). The integration to obtain the total neutral hydrogen column density is performed over the subvolumes extracted from the 3D dust \citet{edenhofer24} map. The orange contours denote six equally spaced bins relative to peak of the total neutral hydrogen column density. These same contours are overlaid atop \hi\ column density images in the right column. The location of the pulsar is shown by the white star symbol. The column density comparison between weaker components is shown in Appendix~\ref{appendix:component_column_density_images}.}
\label{fig:integrated-cdensity}
\end{figure*}

The sightline toward B0823+26 has the least complex \hi\ absorption and emission spectra. The absorption spectrum shows only one velocity component at $4.91$ \kms, which has spin temperature of $62\pm5$ K. The $n_{\rm tot}$ profile shows only one clear dusty structure at 340 pc with $n_{\rm peak}\sim13$ cm$^{-3}$. It is therefore likely that the CNM component seen in \hi\ absorption, which also has a corresponding \hi\ emission feature, is associated with the dust structure at 340 pc. To characterise the morphology of this absorbing structure, we calculate the dust column density image by integrating the sub-volume extracted towards this pulsar over 320$-$360 pc. Using \hi\ emission from the GALFA-HI survey as a guide, we calculate the \hi\ column density image by integrating over the FWHM velocity range of the \hi\ absorption feature (4.91$\pm$2.3 \kms). The two column density images in Figure~\ref{fig:integrated-cdensity} show a well-defined structure with similar morphology, providing confidence that the dusty structure at 340 pc corresponds to the \hi\ structure seen in absorption (and corresponding emission) at about 5 \kms. As the wall of the LB in this direction---shown in green in Figure~\ref{fig:comprehensive_density_profiles}---is at $<200$ pc, this dusty \hi\ structure is located beyond the LB.

B1133+16 has a similarly simple $n_{\rm tot}$ profile, hinting at a single structure in the 3D dust, though the \hi\ emission and absorption show two blended components. The \hi\ absorption spectrum was fitted by \citet{stanimirovic2010} with two CNM components centred at similar radial velocities, $-2$ and $-3.5$ km/s, with the corresponding spin temperature of $\sim30,50$ K, respectively. It is likely that both CNM components, as well as the single WNM \hi\ emission component, are associated with the dusty structure seen at $\sim140$ pc. By integrating the 3D dust sub-volume over the distance range 125$-$175 pc, we find a diffuse dusty structure centred at the pulsar position in Figure~\ref{fig:integrated-cdensity}. The associated \hi\ emission integrated over the velocity range of $-3.5\pm3.6$ reveals a structure at the location of the pulsar sightline. While the dust structure appears diffuse, the \hi\ structure has notable filamentary sub-structure. The agreement between the $n_{\rm tot}$ profile from the dust maps LB structure from the \citet{oneill2024} model in Figure~\ref{fig:mean_density_profiles} suggests that the dusty structure belongs to the LB wall, implying that the blended CNM features seen in \hi\ absorption and clumpy structure in \hi\ emission are also associated with the wall of the LB.  

In contrast to B0823+26 and B1133+16, the \hi\ absorption and especially emission spectra of B1929+10 contain multiple Gaussian components, demonstrating that both WNM and CNM are observed along the line of sight. This sightline is close to the Galactic plane, allowing us to use estimates of the kinematic distance to place some constraints on which dusty structures correspond to individual \hi\ components. The \hi\ absorption spectrum for B1929+10 has three fitted Gaussian components with centroid velocities of $-$0.7 \kms, 4.8 \kms, and 8.4 \kms. We use the Monte Carlo method from a kinematic distance converter tool \citep{wenger2018}\footnote{\url{https://www.treywenger.com/kd/}}, which incorporates the \citet{reid2014} rotation curve and updated solar motion parameters to return more robust kinematic distance uncertainties. Using the velocity centroids from \cite{stanimirovic2010} and the known Galactic latitude and longitude of the pulsar, we attempt to correlate the distances at which peaks in the $n_{\rm tot}$ profile occur with the kinematic distances for each CNM cloud observed along the line of sight in \hi\ absorption.  

The kinematic distance estimates suggest that these components are at distances of 30 pc, 330 pc, and 500 pc for components 1, 2, and 3, respectively. The kinematic distance ambiguity (KDA) within the Sun's orbit, where a given local standard of rest velocity could correspond to a near or far distance, results in kinematic distances that are highly uncertain. The KDA coupled with deviations from circular motion drive our uncertainties to range between 50$-$100\%. Nevertheless, the  
$n_{\rm tot}$ density profile in the direction of B1929+10 shows three peaks at distances of 89 pc, 213 pc, and 278 pc, respectively, and tentatively agree with the returned kinematic distance estimates for the absorbing structures in the \hi\ absorption profile. For example, component 2 with the velocity centroid at 4.8 \kms\ has the highest optical depth (0.120), while the most prominent dust feature in the $n_{\rm tot}$ density profile has the highest density (1.8 cm$^{-3}$). While we cannot unequivocally associate the density peaks with specific CNM components due the large uncertainties in kinematic distances, it is likely that CNM Components 1, 2, and 3 are associated with the peaks at distances of 89 pc, 213 pc, and 278 pc, respectively, when also considering the peak optical depth. Furthermore, we find the LB structure identified by \citet{oneill2024} is associated with the nearest density peak whose kinematic distance is most consistent with the wall of the LB. It is therefore likely that the other two CNM components are located beyond the LB.

The total neutral hydrogen column density images in Appendix~\ref{appendix:component_column_density_images} of the closest and the most distant dust structures towards B1929+10 show very diffuse structure, while the middle (strongest) one in Figure~\ref{fig:integrated-cdensity} shows interesting sub-structure with several higher-density clumps around the pulsar position. The \hi\ column density images for all three components do not show a clear correlation between the dust structures, likely stemming from the complicated sightlines near the Galactic Plane and velocity blending of components centred around LSR velocities of 0 \kms. 

The \hi\ absorption and emission spectra of B2016+28 contain multiple velocity structures which have significant velocity overlap, complicating the decomposition of individual components and radiative transfer calculations and resulting in large uncertainties for the spin temperatures of individual components. The \hi\ absorption spectrum was fitted by 
five Gaussian components all very close in velocity and with peak optical depth ranging from 0.6 to 1.3 \citep{stanimirovic2010}. 

As shown in Figure~\ref{fig:mean_density_profiles}, the dust inferred $n_{\rm tot}$ density profile contains at least five components that are well separated in distance between $\sim200$ and $>800$ pc. By being at a Galactic latitude of $-3.9^{\circ}$, we again use the kinematic distances as a rough distance constraint on the CNM features and correlate with the density structures. The five \hi\ absorption components have kinematic distances ranging from 190 pc to 1.27 kpc, with large uncertainties. The dust-inferred $n_{\rm tot}$ profile also has five components with peaks being within 100$-$200 pc from the kinematic distance estimates for \hi\ absorption components. While this is encouraging, the velocity components with the highest \hi\ optical depth (3.7 \kms\ and 13.3 \kms) seem to correspond to dusty structures with peak $n_{\rm tot}$ values of $\sim1$ cm$^{-3}$ and $\sim9$ cm$^{-3}$, respectively. Also, the dusty structure with the highest peak $n_{\rm tot}$ corresponds to one of CNM structures with relatively low optical depth ($\sim0.8$). This shows the the correspondence between \hi\ absorption and dusty structures is not easy to disentangle for busy sightlines near the Galactic Plane. Our strategy of calculating \hi\ column density images over the velocity ranges of individual CNM components and comparing their morphology with the dust column density images also did not result in simple matches. However, the second, fourth, and fifth components correspond to clumpy structures in the total neutral hydrogen column density images in the left columns of Figure~\ref{fig:appendix_integrated-cdensity}. Nonetheless, we conclude that this pulsar is probing a direction that is too complex to disentangle and match to dusty structures. This is consistent with the large uncertainties in spin temperature estimates stemming from fitting Gaussians to blended CNM and WNM components. Finally, we find that the closest dusty structure is located at the wall of the LB, according to the \citet{oneill2024} model. The large optical depth of this CNM feature suggests that the structure of the LB wall appreciably contributes to the \hi\ absorption in this direction. 

\subsection{Lower Limits on the Thermal Pressure}\label{subsec:results_thermal_pressure}

Using our connections between individual \hi\ absorption components and density structures established in Section~\ref{subsec:correlate_abs_with_ntot} and summarised in Tables~\ref{tab:cnm_derived_properties} and ~\ref{tab:dust_derived_properties}, along with the DM values in Table~\ref{tab:pulsar_properties}, we calculate a lower limit on the thermal pressure and upper limit on the ionisation fraction. For the thermal pressure, we use two methods. Firstly, we assume the CNM is an ideal gas and use the $\langle n_{peak} \rangle$ density as an estimate of the particle volume density within the structure. For each density component, we multiply this density by the corresponding $T_{\rm s}$, which at the densities of the CNM we assume is fully thermalised by collisions and thus a proxy for kinetic temperature, to obtain an estimate for the thermal pressure in units of cm$^{-3}$ K. These values are shown under the $P/k$ column in Table 4. The uncertainties are statistically propagated from the systematic errors on $\langle n_{peak} \rangle$ and $T_{\rm s}$. We note that since the dust data are interpolated and smoothed onto a 1 pc$^{-3}$ voxel, the peak density values we report are effectively lower limits on the true density in any given LOS and, thus, our thermal pressure estimates are also strictly lower limits. We call this approach the ``direct" thermal pressure.

Our second method for estimating the thermal pressure takes advantage of the 3D dust data constraining the LOS extent of the absorbing structures. From $n_{\rm tot}$ density profiles shown in Figures~\ref{fig:mean_density_profiles}, we measure the FWHM of the LOS extent for each of nine structures and show these values in Table~\ref{tab:dust_derived_properties}. These range between 10–40 pc. As \citet{stanimirovic2010} measured the CNM column density for each absorption component, as well as the ratio of the CNM mass fraction (the ratio of the CNM column density to the CNM+WNM \hi\ column density along each LOS), we estimate the total volume density of \hi\ ($<n_{HI}>$) for each structure by dividing the CNM column density by the mass fraction. The result is then divided by the LOS extent to obtain an estimate of the average \hi\ volume density, $\langle n_{\rm HI} \rangle$. Our primary assumptions are that our connections between the CNM components in velocity and dust structures are correct, which is more uncertain for the complex sightlines near the Galactic Plane, and the \hi\ and dust are well-mixed to trace similar 3D depths. We estimate the thermal pressure $P'/k$ in the \hi\ by multiplying $\langle n_{\rm HI} \rangle$ with the associated $T_{\rm s}$ values. As this estimates considers only the neutral hydrogen, this is again a lower limit. We call this second method the ``geometric" thermal pressure. Here, the lower bounds are primarily limited by high uncertainties in the $T_{\rm s}$ values. We list these estimates of the neutral hydrogen density and resulting thermal pressure in Table~\ref{tab:dust_derived_properties}. The uncertainties are statistically propagated from the systemic errors on $N_{\rm HI,CNM}$ and the FWHM span.

\begin{figure}
    \centering
    \includegraphics[width=0.5\textwidth]{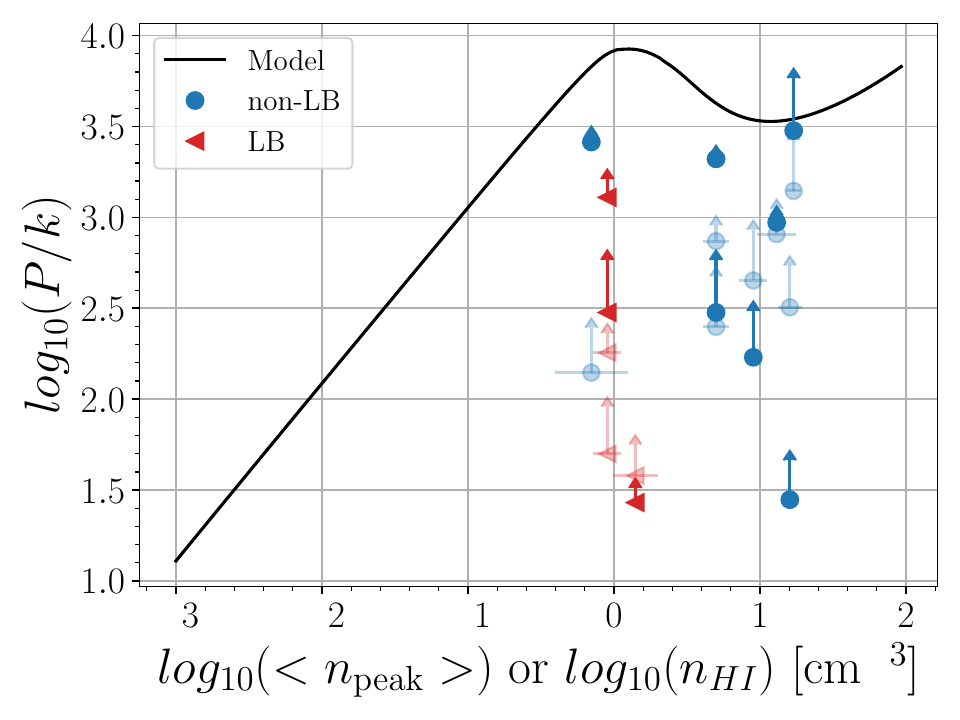}
    \caption{Comparison between the lower limits of our estimates for thermal pressure using the direct (transparent) and geometric (opaque) methods for each dust and CNM component along the LOS towards each pulsar as function of mean peak total neutral hydrogen computed from the 3D dust maps (contained from the direct approach) or the mean HI volume density (constrained from the geometric approach). The black solid line shows the equilibrium curve expected for the Solar neighbourhood from \citet{wolfire2003}. The length of the lower limit arrows represent the relative statistical uncertainty.}
    \label{fig:thermal_pressure}
\end{figure}

Figure~\ref{fig:thermal_pressure} makes a comparison between the equilibrium curve in the Solar neighbourhood from \citet{wolfire2003} and our lower limits on the thermal pressure, distinguishing between structures inside and outside of the LB. Most of the structures have a thermal pressure in the 20--700 K cm$^{-3}$ range using the direct approach. This is a factor of a few to $\sim$100 times lower than what is expected for thermal equilibrium based on \cite{wolfire2003}. Only the dusty structure located at $\sim760$ pc in the direction of B2016+28 has a thermal pressure of $\sim1400$ K cm$^{-3}$, which is a factor of few lower than what is expected for the Solar neighbourhood thermal equilibrium conditions. Our geometric approach to estimating thermal pressure in \hi\ along the LOS yields lower limits in better alignment with the theoretical expectations for the solar neighborhood. For example, the structure toward B2016+28 now has a thermal pressure of $\sim$3000 K cm$^{-3}$, though with large error bars due to the substantial uncertainties in estimating $T_{\rm s}$ from the complex spectral profiles typical of Galactic Plane sightlines. The geometric method also raises the lower limits on thermal pressure to within a factor of a few of the \citet{wolfire2003} predictions for all LOS structures toward B1929+10 and B0823+26. This improvement shows that pencil-beam \hi\ absorption measurements, when combined with lower-resolution 3D dust maps to constrain the depth of absorbing structures along the LOS, can yield reliable lower limits on the ISM thermal pressure. We also note that the $\langle n_{\rm HI} \rangle$ densities are typically higher than the peak $\langle n_{\rm peak} \rangle$ estimated from 3D dust data alone, emphasising again that 3D dust cubes suffer from averaging over $\sim1$ pc$^3$, while the \hi\ absorption detects CNM that is likely clumpy on scales smaller than 1 pc.

\subsection{Upper Limits on the Ionisation Fraction}\label{subsec:results_ionising_fraction}

The ionisation fraction can be used to categorise the \hi\ structures as either WNM, CNM, or perhaps a combination of both. Due to a combination of embedded and external sources of extreme ultraviolet, X-ray, and cosmic ray radiation, the WNM near the Solar neighbourhood is expected to have a ionisation fraction of $\sim1.7\times10^{-2}$, while the CNM is expected to be less ionised at $\sim4\times10^{-4}$ \citep{draine2011_txtbk}. The ionisation fraction is defined as
\begin{equation}
    \langle x_e \rangle  = \langle n_e \rangle/\langle n_{\rm tot} \rangle, 
\end{equation}
where $\langle n_e \rangle$ is the mean electron density along the LOS. We first estimate $\langle n_e \rangle$ by dividing the DM values listed in Table~\ref{tab:pulsar_properties} by the distance to each pulsar. We note again that structures within the 3D dust maps are interpolated to 1 pc$^{-3}$ voxels, effectively washing out sub-parsec density fluctuations, resulting in a lower limit on the true $\langle n_{\rm tot} \rangle$ for a given LOS; thus, our estimates for $\langle x_e \rangle$ should be taken as an upper limit.

Table~\ref{tab:los_properties} summarises our upper limit on the mean ionisation fraction along the LOS towards each pulsar. Both of our $\langle n_e \rangle$ and $\langle n_{\rm tot} \rangle$ values are consistent with the implied electron densities from \citet{jenkins2013} at \hi\ volume densities of $\sim0.5$ cm$^{-3}$ within a few hundred parsecs of the Sun. We note two points: (1) that $\langle n_e \rangle$ values from DMs, as they preferentially intersect highly ionised regions around pulsars, can be inflated relative to sightlines not directed towards pulsars; and (2) that a direct comparison with the observations of \citet{jenkins2013} is tenuous since the ratio of neutral species \ari~and \oi~column densities was used as a proxy for the fractional ionisation of \hi\ only. Nevertheless, our upper limits are consistent with these measurements. Unfortunately, the limited physical resolution of the 3D maps prevents us from robustly testing the ultimate conclusion of \citet{jenkins2013}---that the systematic deficiency of \ari~results from ionisation caused by supernova explosions, whose effects on the WNM persist long after the remnants are visible in the X-rays. 

Table~\ref{tab:los_properties} shows upper limits on the fractional ionisation of $\sim$1\% are associated with the lowest latitude Galactic pulsars, indicating sightlines through the Galactic Plane can provide better constraints. Though one also needs to account for the bias introduced by focusing on sightlines that inherently intersect regions of high ionisation towards pulsars. Nevertheless, our upper limits towards low Galactic latitudes are consistent with $\langle x_e \rangle\sim0.003$ values measured for diffuse clouds in the Galactic Plane \citep{welty2003}. This is higher than what would be expected from photoionisation from carbon and other heavy elements, suggesting the $\langle x_e \rangle$ is overestimated. Our upper limits are in better agreement with the theoretical predictions for $\langle x_e \rangle$ from \citet{Godard2024} using Paris-Durham shock codes, which range between 0.01 and 0.03 when a $\sim$200 km s$^{-1}$ shock interacts with the surrounding ISM. Ultimately, a larger sample of sightlines towards pulsars in the Galactic plane is needed to make robust comparisons with published observational and theoretical estimates for the fractional ionisation fraction. 

\begin{table*}
    \centering
    \caption{Derived CNM Properties from \hi\ Absorption}
    \begin{tabular}{ccccccc}
    \hline\hline
         Pulsar &  $\tau_0$&  $v_0$ (LSR)&  $T_s$&  $N_{\hi,\rm CNM}$& $\Delta v$ (LSR) &  $\frac{N(\hi)_{CNM}}{N(\hi)_{tot}}$\\
 Component & & [km s$^{-1}$]& [K]&  [$10^{20}$cm $^{-2}$]& [km s$^{-1}$] &\\
 \hline
         B0823+26&  $0.26 \pm 0.01$&  $4.91 \pm 0.02$&  $62 \pm 5$&  $0.72 \pm 0.02$& 3.8 to 6.1 &0.14\\
         B1133+16&  &  &  &  & $-$5.3 to $-$1.25 & 0.16\\
         1&  $0.16 \pm 0.02$&  $-2.91 \pm 0.03$&  $27 \pm 5$&  $0.15\pm0.01$& \\
         2&  $0.16 \pm 0.01$&  $-3.5 \pm 0.1$&  $48 \pm 6$& $0.53\pm0.03$& \\
         B1929+10&  &  &  &  & $-3.4$ to 10.4 & 0.07\\
         1&  $0.017 \pm 0.002$&  $-0.7\pm 0.6$&  $199\pm25$&  $0.35\pm0.04$& \\
         2&  $0.120 \pm 0.005$&  $4.8 \pm 0.1$&  $148\pm4$&  $1.00\pm0.04$& \\
         3&  $0.076 \pm 0.007$&  $8.4 \pm 0.2$&  $206\pm7$&  $1.15\pm0.06$& \\
         B2016+28&  &  &  &  & 2.5 to 14.3 & 0.39\\
 1& $1.12\pm 0.03$& $3.7 \pm 0.03$& $50\pm30$& $2.8\pm0.3$
&\\
 2& $0.67 \pm 0.04$& $5.97 \pm 0.04$& $50\pm30$& $0.9\pm0.1$
&\\
 3& $0.81 \pm 0.05$& $9.16 \pm 0.06$& $80\pm50$& $13.3\pm1.5$&\\
 4& $0.57 \pm 0.05$& $9.70 \pm 0.04$& $50\pm30$& $1.7\pm0.5$&\\
 5& $1.30 \pm 0.03$& $13.3\pm 0.01$& $20\pm10$& $0.7\pm0.1$&\\
 \hline
    \end{tabular}
    \begin{minipage}{\linewidth}
    \vspace{1mm}
        NOTE: Columns (2)$-$(7) respectively summarise the peak optical depth, velocity centroid, spin temperature for each \hi\ absorption component, the span in velocity encompassing the FWHM of the CNM components, the CNM \hi\ column density, and the CNM mass fraction for the sightline measured in \citet{stanimirovic2010}.
    \end{minipage}
    \label{tab:cnm_derived_properties}
\end{table*}

\begin{table*}
    \centering
    \caption{Derived Dust Structure Properties from the $n_{\rm tot}$ Profiles (this work).}
    \begin{tabular}{ccccccccc}
    \hline\hline
         Pulsar &  FWHM span&  Peak&  Kin. distance & $|z|$  & $\langle n_{peak} \rangle$&   $\frac{P}{k}$&$\langle n_{HI}\rangle$ &$\frac{P}{k}'$\\
Component & [pc]& [pc]& [pc]& [pc] & [cm$^{-3}$]& [K cm$^{-3}$]&[cm$^{-3}$] &[K cm$^{-3}$]\\\hline
         B0823+26&  $11\pm3$ &  $340.0\pm0.9$ &  n/a & 179 & $13\pm4$ &  $810\pm260$ & $15.2\pm0.4$ &$939\pm80$\\
         B1133+16&  &  &  &  &  &&\\
         1&  $30\pm11$&  $138\pm5$&  n/a & 129 &  $1.4\pm0.5$
&  $38\pm15$& $1.01\pm0.07$& $27\pm5$\\
         2&  n/a&  n/a&  n/a&  n/a&  n/a& n/a&n/a\\
         B1929+10&  &  &  &  &  &&\\
         1& $18\pm7$&  $89\pm6$&  $30^{+330}_{ -20}$& 6 & $0.9\pm0.2$ &  $180\pm45$
& $9\pm1$&$1791\pm304$\\
         2&  $33\pm7$&  $213\pm3$&  $330^{+230}_{-320}$& 14  &$5\pm1$&  $740\pm150$& $14.0\pm0.5$&$2076\pm100$\\
         3&  $30\pm14$&  $280\pm10$&  $500^{+380}_{-290}$& 19  & $0.7\pm0.4$&  $140\pm80$ & $17.1\pm0.9$&$2656\pm228$\\
         B2016+28&  &  &  &  &  &&\\
 1& $37\pm8$& $205\pm6$& $190^{+730}_{-180}$& 14 & $0.9\pm0.2$& $50\pm30$ & $6.3\pm0.7$&$314\pm192$\\
 2& $22\pm8$& $405\pm5$& $520^{+600}_{-510}$& 28 & $9\pm2$ &  $450\pm290$ & $3.4 \pm 0.4$ & $170\pm104$\\
 3& $28\pm7$& $763\pm3$& $720^{+680}_{-620}$& 53  & $17\pm2$ & $1400\pm900$& $39\pm1$&$3157\pm2005$\\
 4& $26\pm8$& $835\pm7$& $1140^{+340}_{-1030}$& 58 & $5\pm1$& $250\pm160$ & $5.4\pm0.6$&$272\pm182$\\
 5& $42\pm8$& $942\pm8$& $1270^{+680}_{-840}$ & 65 & $16\pm3$& $320\pm170$& $1.4\pm0.2$&$28\pm14$\\
 \hline
    \end{tabular}
    \begin{minipage}{\linewidth}
    \vspace{1mm}
    Columns (2)$-$(9) respectively provide the FWHM of each density component along the LOS, the location of the peak density ($d_{\rm peak}$), the kinematic distance (see Section~\ref{subsec:correlate_abs_with_ntot}), the vertical distance from the Galactic Plane ($|z|=d_{\rm peak}\sin{b}$), the peak density value averaged over all 12 realisations, the lower limit on the geometric thermal pressure, the average mean \hi\ volume density, and the geometric thermal pressure.
    \end{minipage}
    \label{tab:dust_derived_properties}
\end{table*}

\section{Discussion}\label{sec:discussion}

\subsection{TSAS towards B1929+10 is not associated with the Local Bubble wall}\label{subsec:tsas_b1929}

\citet{stanimirovic2010} found persistent variations in the \hi\ absorption profiles towards B1929+10 over four epochs. Based on the presence of Na I absorption from \citet{genova1997} towards two nearby ($<3^{\circ}$ from the pulsar sightline) stars---HD 178125 at a distance of 173 pc and HD 180555 at a distance of 106 pc with central velocities at 5.9$\pm$1.0 \kms\ and 5.6$\pm$1.0 \kms, respectively---they suggested an interstellar cloud either within or on the LB wall is responsible for these absorption features, indicating the \hi\ absorption variability is linked to TSAS on AU scales. However, the most pronounced variation was seen in our Component 2 centred at 4.8 \kms, which we associate with a dusty structure along the line of sight located at a distance of $\sim215$ pc. This is more than 100 pc more distant than the LB structure identified by \citet{oneill2024} (see Figure~\ref{fig:mean_density_profiles}). Interestingly, this LB structure is located within the upper bound of 106 pc distance suggested by the Na I absorption towards HD 180555. However, the prominence of the more distant density feature, the offset between the stellar and pulsar sightlines, and the absence of a corresponding structure in an independent LB model based on the same dust map all suggest that most of the TSAS fluctuations observed by \citet{stanimirovic2010} are not associated with a cloud within or on the LB wall, but rather with another more distant structure along the same line of sight. We do not see any indication of small-scale structure corresponding to TSAS, though this is feasibly due to the low resolution of the dust maps. Ultimately, the limitation of physical resolution and lack of data within the inner 69 pc of the Sun preclude us from definitively concluding that the cloud causing the TSAS is located outside the LB. Nevertheless, our comparison demonstrates promise for identifying structures within 3D dust maps along sightlines with observed TSAS to better understand the properties of the LB wall and the multi-phase ISM that lies beyond. 

\subsection{\hi\ Absorption Inside and Outside the LB}\label{subsec:LB_absorption}

Our analysis shows that three CNM features seen in \hi\ absorption are feasibly associated with the LB wall when compared with the \citet{oneill2024} model: the primary component towards B1133+16, Component 1 towards B1929+10, and Component 1 towards B2016+28. The lower limit on $<n_{\rm peak}>$ from the 3D dust maps reveal systematically lower values than those made outside the LB, consistent with the notion that the Local Bubble is occupied by a warm/hot ionised medium traced in X-rays that easily evaporates cloud atomic clouds \citep{snowden1998}. Regardless of the method used, our lower limits on the thermal pressure do not show any systematic differences between components we associate as being inside and outside the Local Bubble.   

While there are detections of CNM clouds associated with the Local Bubble Wall \citep{rybarczyk2024}, the majority of the CNM structures seen in \hi\ absorption towards these pulsars are seemingly associated with dusty structures at distances of 200 pc to 500 pc. Our $n_{\rm tot}$ profiles demonstrate that this is especially true for sightlines near the Galactic Plane. However, we note the direct association between CNM detected through \hi\ absorption and distant dusty structures depends on the uniqueness of the kinematic distance estimates, making the matching process highly uncertain for the lower latitude sightlines. Nevertheless, the most prominent density features along these LOS do not correspond to the structure of the Local Bubble Wall, again aligning with the view that the Local Bubble is generally devoid of CNM clouds---with some notable exceptions like the Leo Cold Cloud \citep{verschuur1969, heiles2003b}

The sightline towards B0823+26 demonstrates that the association between \hi\ absorbing and dust structures is relatively unique and reasonable towards higher Galactic Latitudes. Complex structure in the sightlines and the distance ambiguity inherent to kinematic distance estimates make it extremely challenging to match CNM clouds to 3D dust structures. Similarly, the velocity crowding in \hi\ emission makes it complex to match \hi\ structures traced by emission spectra with 3D dust structures. Our study establishes that future work looking to correlate the atomic ISM with the 3D structures in dust maps should focus on correlating simple ($<3$ components) absorption spectra towards intermediate to high ($|b|\gtrsim10^{\circ}$) Galactic latitudes.

\subsection{The Thermal Pressure of the Local ISM}\label{subsec:thermal_pressure}

We place a lower limit of the thermal pressure within the total neutral hydrogen and \hi\ towards each pulsar through the direct and geometric approaches discussed in Section~\ref{subsec:results_thermal_pressure}. The lower limits imposed by the direct approach are effectively impractical due to the resolution differences between the interpolated dust maps and pencil-beam sightlines probed by the \hi\ absorption observations. The geometric approach, however, utilises the dust maps to constrain only the depth of a structure associated with a CNM component provided by the Gaussian decomposition in \citet{stanimirovic2010}, resulting in more reasonable lower limits towards some pulsars---noting again that these calculations depend on the assumptions that the \hi\ and dust are well-mixed and individual CNM components are successfully matched to dusty structures along the LOS. For example, the lower limits placed on all components towards B1929+10 and the most prominent density peak (Component 3) towards B2016+28 are within less than a factor of two of $P/k=3800$ cm$^{-3}$ K---the mean gas pressure in the Solar neighbourhood \citep{jenkins2011}. The PDF of thermal pressure from \citet{jenkins2011} shows a tail towards higher values such that less than 1\% of observed clouds have thermal pressures greater than 10000 cm$^{-3}$ K, placing an effective upper limit on the thermal pressure in the atomic gas within the Solar neighbourhood. The high-pressure regions, found close to massive stars and likely due to shocks and stellar winds, have been confirmed recently by \citet{jenkins2021}. In regards to \hi\ specifically, numerical and analytical work (e.g., \citealt{wolfire2003, bialy2019}) predicted a minimum pressure for CNM of 3000 K cm$^{-3}$ with the WNM able to exist in pressure equilibrium up to $\sim$6500 K cm$^{-3}$ for the Solar neighbourhood. The presence of \hi\ emission in our sightlines, which predominately traces the WNM due to low $T_{\rm s}$ for the CNM, indicates the thermal pressure in the \hi\ must lie within this range. Thus, the lower constraints on the thermal pressure in \hi\ suggest our sightlines nearest to the Galactic Plane intersect environments typical of the Solar neighbourhood. 

\begin{figure}
    \centering
    \includegraphics[width=0.5\textwidth]{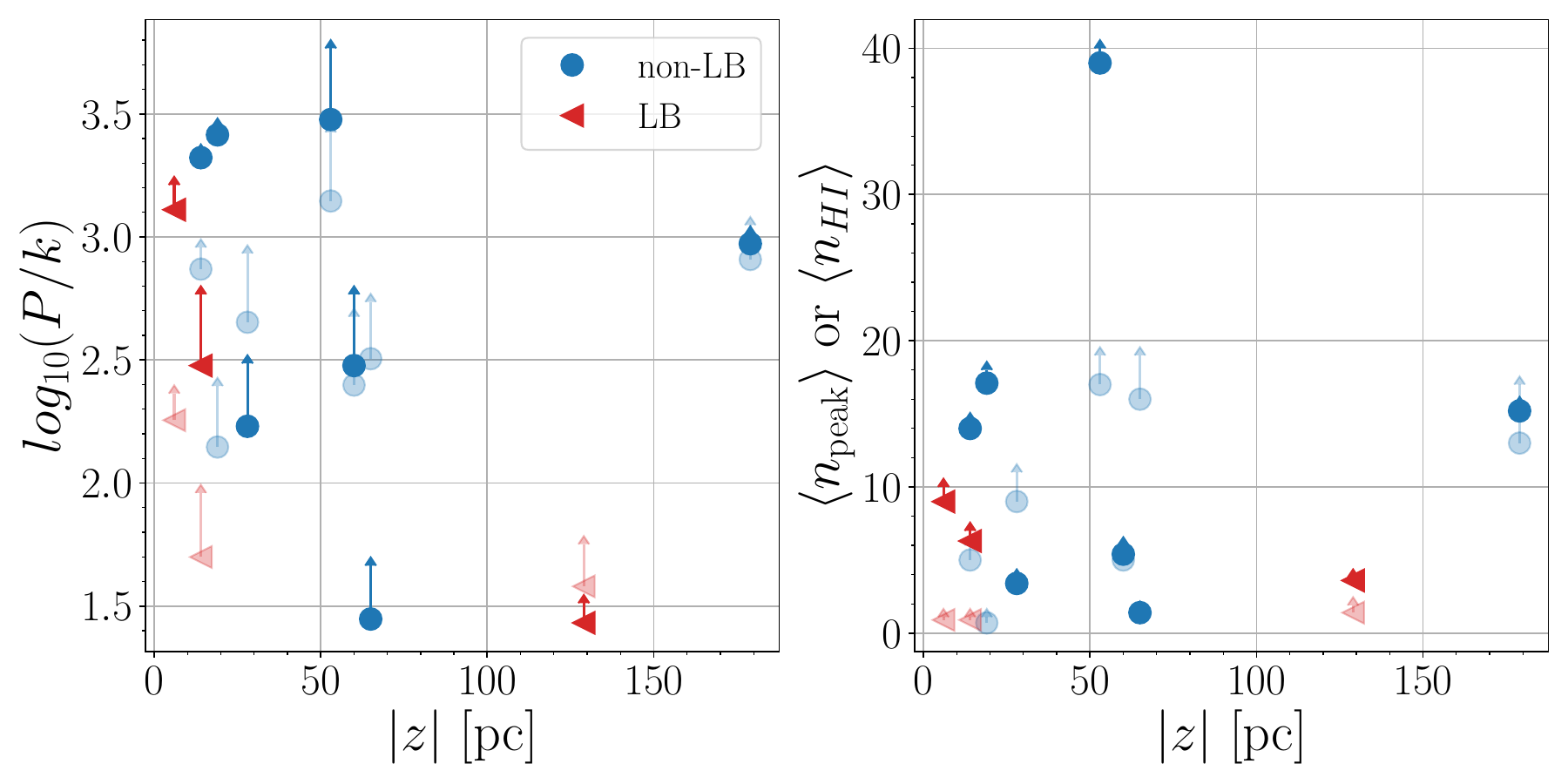}
    \caption{Left: The lower limits for the direct (transparent) and geometric (opaque) thermal pressure for each dust and CNM component along the LOS towards each pulsar as a function of vertical distance from the Galactic Plane. Right: the lower limits for the direct and geometric volume densities, respectively constraining the total neutral hydrogen and \hi\ volume densities, as a function of vertical distance from the Galactic Plane. The length of the lower limit arrows represent the relative statistical uncertainty.}
    \label{fig:pressure_density_absz}
\end{figure}

On the other hand, the lower limits on the thermal pressure towards pulsars at high Galactic latitudes are systematically offset from the those towards pulsars that reside within the Galactic Plane, with no clear difference between the direct and geometric approaches to constraining the thermal pressure. The distance to the density peaks towards B0823+26 and B1133+16 and $b$~values imply vertical extents ($|z|$) of $\sim$210 pc and $\sim$360 pc for these isolated CNM clouds, or four times and seven times the thickness of the CNM disk for clouds with $\tau>0.1$ \citep{rybarczyk2024}. Figure~\ref{fig:pressure_density_absz} shows the lower limits for the thermal pressure and volume densities from the direct and geometric approaches as a function of vertical distance from the Galactic Plane. Regardless of the direct or geometric approach, the lower limits on the thermal pressure and volume density are, in general, inversely proportional the components vertical distance from the Galactic Plane. These systematically reduced lower limits for thermal pressure and volume density along sightlines outside of the Galactic Plane are consistent with the picture that, outside of the thin CNM disk, the total pressure decreases due to the reduced weight of the ISM. 

While our sightlines seemingly probe typical conditions in the local ISM, clouds with extremely high pressures do exist within the Local Bubble, likely resulting from the warm cloud collisions. For example, the Leo Cold Cloud at a distance of 11 pc--24 pc \citep{meyer2006, peek2011} is located inside the Local Bubble. \citet{heiles2003b} estimate its kinetic temperature to be 20 K based on \hi\ absorption detections. Its total neutral hydrogen column density of $3\times 10^{19}$ cm$^{-2}$ suggests it has a sheet-like geometry and a line-of-sight thickness of $<0.1$ pc. Analysis of UV spectra towards two background stars by \citet{meyer2012} imply a striking mean thermal pressure of 60000 cm$^{-3}$K at an inferred \hi\ volume density of 3000 cm$^{-3}$, potentially caused by the colliding flows of warm gas. Unfortunately, the Leo Cold Cloud falls within the inner 69 pc that is not available in the \citet{edenhofer24} maps. Future work will focus on systematically applying our geometric approach for placing a lower bound on the thermal pressure to a large sample of nearby molecular clouds with \hi\ absorption \citep{stanimirovic2014, Nguyen2019}, constraining the geometry---e.g., filament vs. sheet-like---and comparing their environments to diagnose the mechanisms producing structures with remarkable pressures like the Leo Cold Cloud.



\section{Conclusions \& Future Work}\label{sec:conclusions}

We present total neutral hydrogen density profiles along the line of sight to four local pulsars---B0823+26, B1133+16, B1929+10, and B2016+28---and draw correlations to the corresponding absorption and emission spectra previously measured by \citet{stanimirovic2010}.  We compare the column density images of \hi and total neutral hydrogen density towards each pulsar to draw connections with the dusty and gaseous structures. We further associate the structures along each pulsar's line of sight to those identified to be part of the Local Bubble by the model presented in \cite{oneill2024} and \hi\ absorption spectra from \citet{stanimirovic2010}. Through our measurements of the peak hydrogen and average \hi\ volume densities for each component along the line of sight to each pulsar, we place lower and upper limits on the thermal pressure and lower limits on ionisation fraction, respectively, to work towards a quantitative characterisation of the gas as either CNM or WNM. Our main findings are summarised below: 
\begin{itemize}
    \item The density profile of B0823+26 has a single high-density (10 cm$^{-3}$) peak near 350 pc, matching the sole Gaussian velocity component (4.91 km s$^{-1}$) in the associated \hi\ absorption spectra. The total neutral hydrogen column density image taken over the FWHM distance reveals a well-defined clump of gas near the LOS, providing confidence that the density peak and velocity component correspond to a small CNM gas cloud located beyond the LB. Similarly, the density profile of B1133+16 has a single low-density (0.5 cm$^{-3}$) peak near 150 pc. The associated absorbing structure is a diffuse cloud is located in the wall of the LB, and is correlated with the two CNM components ($-$2, $-$3.5 km s$^{-1}$ respectively) of the absorption spectra as well as the single \hi\ component of the emission spectra. 
    \item  Given their proximity to the Galactic plane, B1929+29 and B2016+28 have multiple density peaks along their respective lines of sight. We roughly correlate the first three peaks of the density profile of B1929+10 located at 89, 213, and 278 pc with the first 3 components of the absorption spectra with mean velocities of $-$0.7, 4.8, and 8.4 km s$^{-1}$. We associate the primary density and CNM component lie outside of the Local Bubble, indicating that the TSAS observed in \citet{stanimirovic2010} also occurs beyond the wall of the Local Bubble. The total neutral hydrogen column density taken along the FWHM for each peak shows different levels of diffusivity in the gas surrounding the pulsar, revealing the complexity of the dusty structures in small increments of distance along the line of sight. Meanwhile, the first density peak in the density profile of B2016+28 probably corresponds to an absorbing structure in the LB wall, but the four other peaks are not well-correlated with the four other Gaussian components in the absorption spectra. 
    \item We calculate lower limits of the thermal pressure by using the ideal gas law, employing two different methods to measure the \hi\ volume density: (a) directly measuring the peak density and (b) dividing the FWHM density value of each peak by the CNM fraction as measured by \citep{stanimirovic2010}. The geometric approach (b) yields lower limits closer to the thermal equilibrium curve as set by \citep{wolfire2003} since the absorption spectra are more sensitive to variations in the density on a smaller scale than the three-dimensional dust maps. The upper limits of the ionisation fraction along the line of sight to each pulsar are consistent with the WNM calculations of  \citep{jenkins2013}. Also, our measurements are in agreement with theoretical predictions from \citep{Godard2024}, signifying the possibility of of $\sim$200 \kms\ shock interacting with the ISM along the LOS. 
\newline\newline
Our study shows promise in physically correlating dusty structures via density profiles to the CNM probed by \hi\ absorption spectra along the line of sight towards pulsars with higher Galactic Latitudes. Furthermore, our analysis shows both diffuse WNM and clumpy CNM structures are correlated with \hi\ absorption, with several components lying outside of the LB. Studies that incorporate more LOS samples towards local pulsars will place important constraints on its thermal properties of the LB and elucidate its overall evolution. 

\end{itemize}

\section{Acknowledgments}
 We would like to thank Prof.~Robert Benjamin, Prof.~Shmuel Bialy, and Dr.~Catherine Zucker for providing essential data sets, improving our methodology and fostering discussions that led us to finding new results. Additionally, we thank the anonymous referee whose detailed suggestions improved the clarity and quality of this work.

The three-dimensional visualisation of the data was made possible thanks to the Glue visualisation software, supported under NSF grant numbers OAC-1739657 and CDS\&E:AAG-1908419.

Software: Astropy \citep{astropy2013, astropy2018, astropy2022}; Dustmaps \citep{green2018}; Glue \citep{beumont2015, robitaille2019}; Healpy \citep{zonca2020}; Matplotlib \citep{Hunter2007}; Numpy \citep{harris2020array}.

\appendix
\section{Column Density Images of Individual Components}\label{appendix:component_column_density_images}
In Section~\ref{subsec:correlate_abs_with_ntot}, we compare the morphology of the total total hydrogen and \hi\ column density images integrated over the FWHM centred on the location of the density (velocity) peak for primary total neutral hydrogen (\hi) component towards each pulsar. Figure~\ref{fig:appendix_integrated-cdensity} shows these maps for the weaker components towards B1929+10 and B2016+28. Unlike the primary second component, the column density maps of the first and third components towards B1929+10 show diffuse structure in both density and \hi\ emission. Due to the significant velocity overlap of the fitted \hi\ absorption components, it is difficult to draw clear correlations with the sub-structures in both maps. This demonstrates that physical correlation between density structures in 3D dust maps with \hi\ absorption should focus on high-latitude sightlines with simple ($<$3 components) density profiles. 

\begin{figure*}
    \centering
    \includegraphics[width=0.40\textwidth]{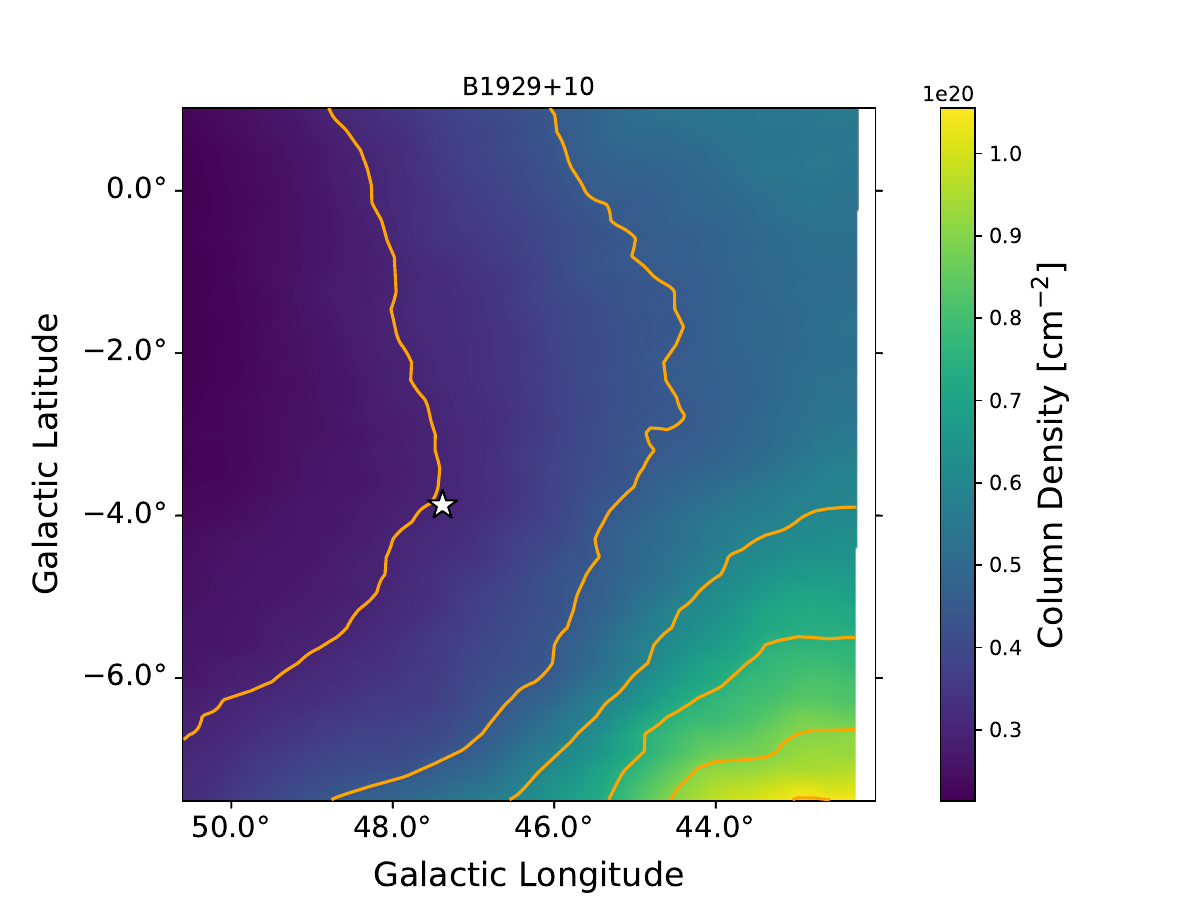}
    \includegraphics[width=0.40\textwidth]{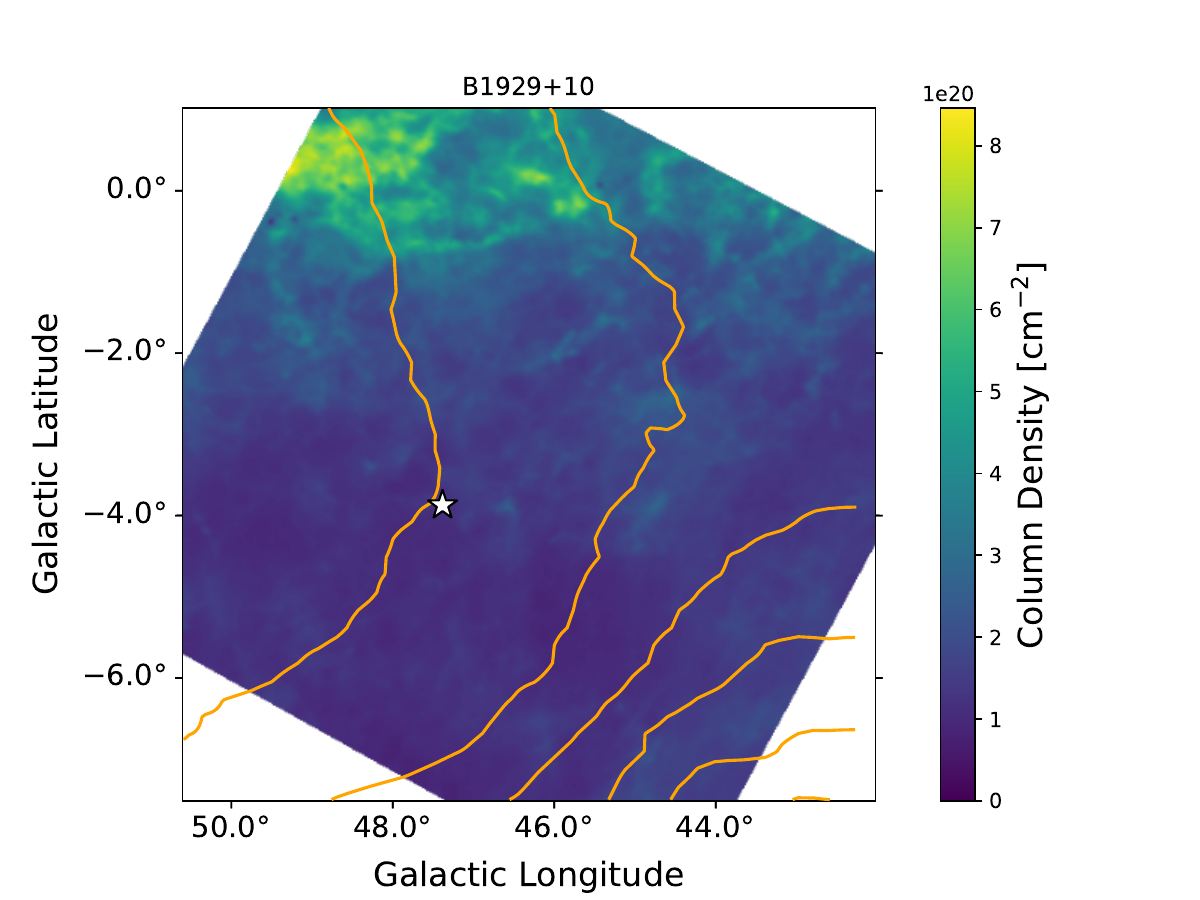}
    \includegraphics[width=0.40\textwidth]{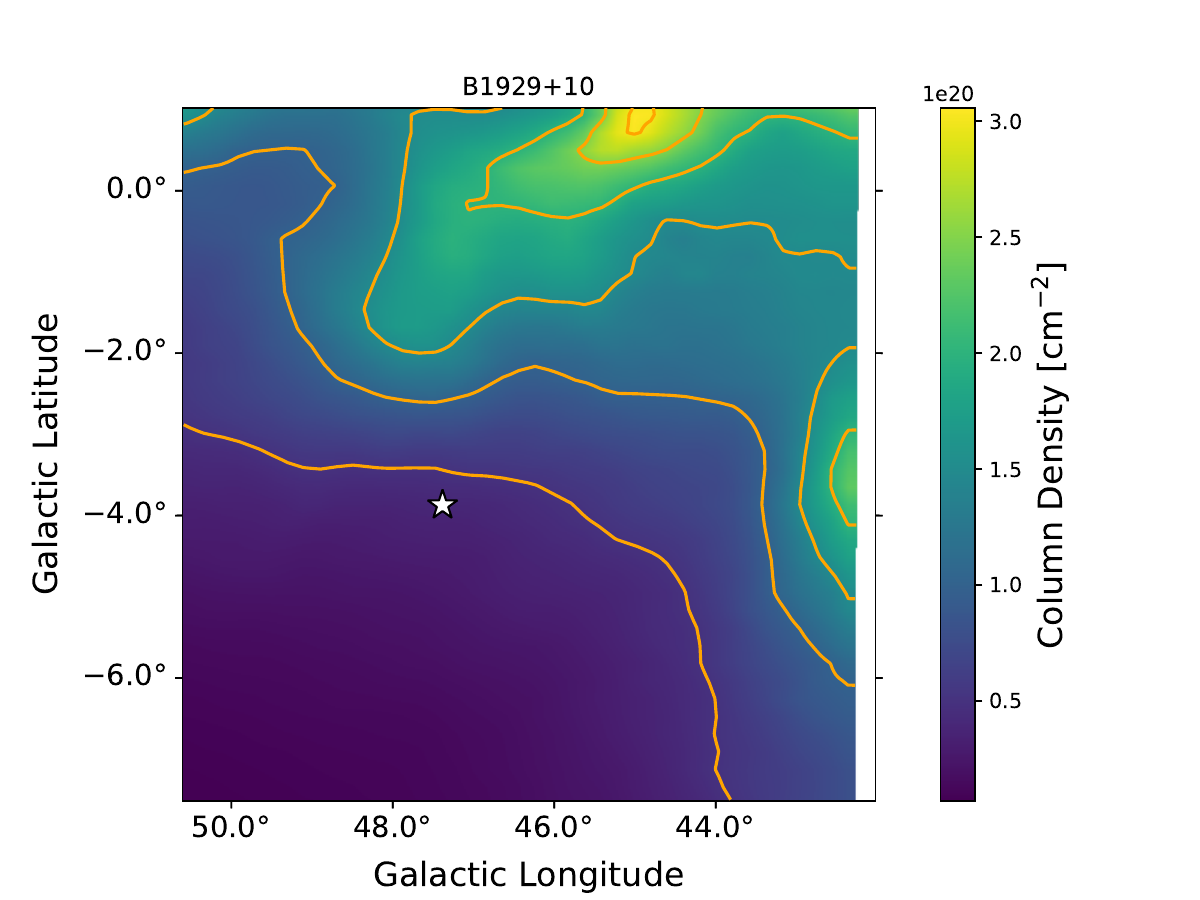}
    \includegraphics[width=0.40\textwidth]{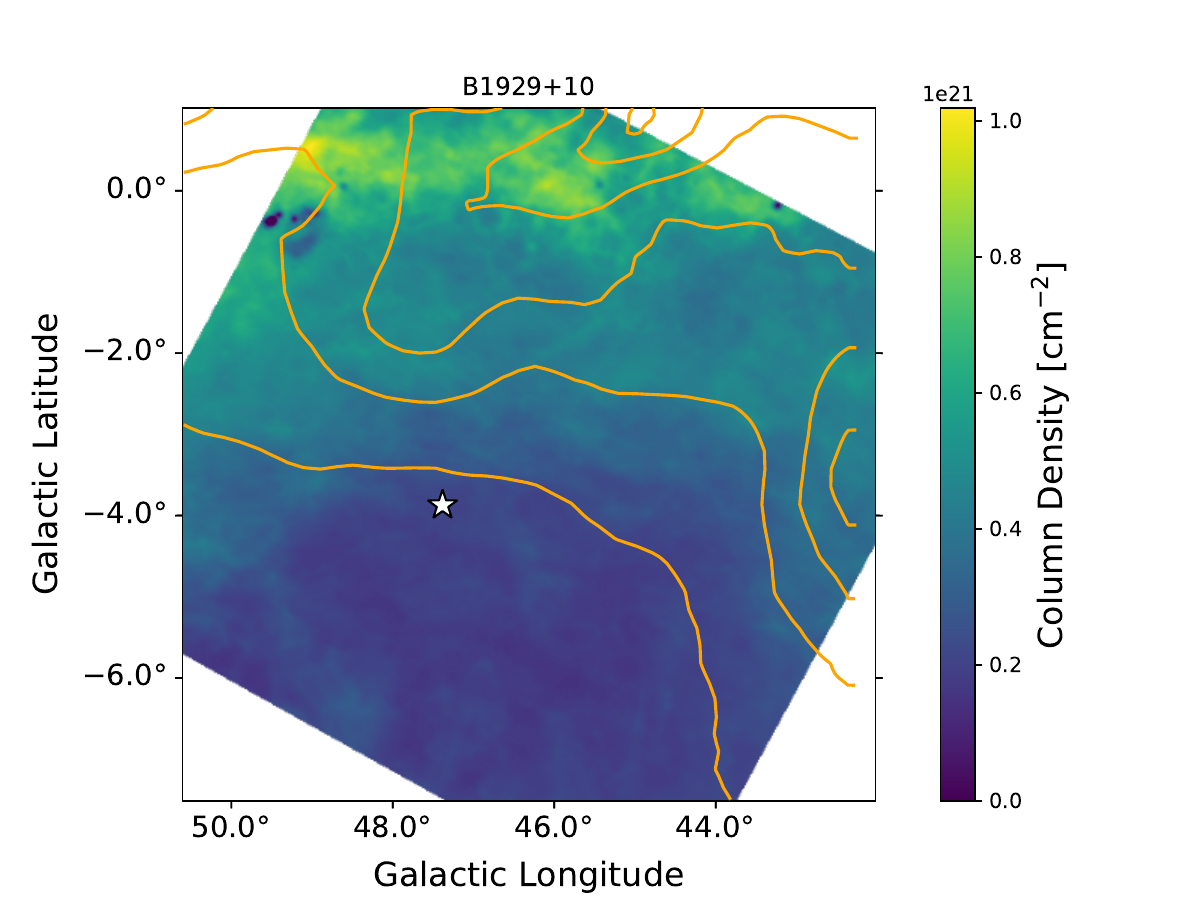}
    \includegraphics[width=0.40\textwidth]{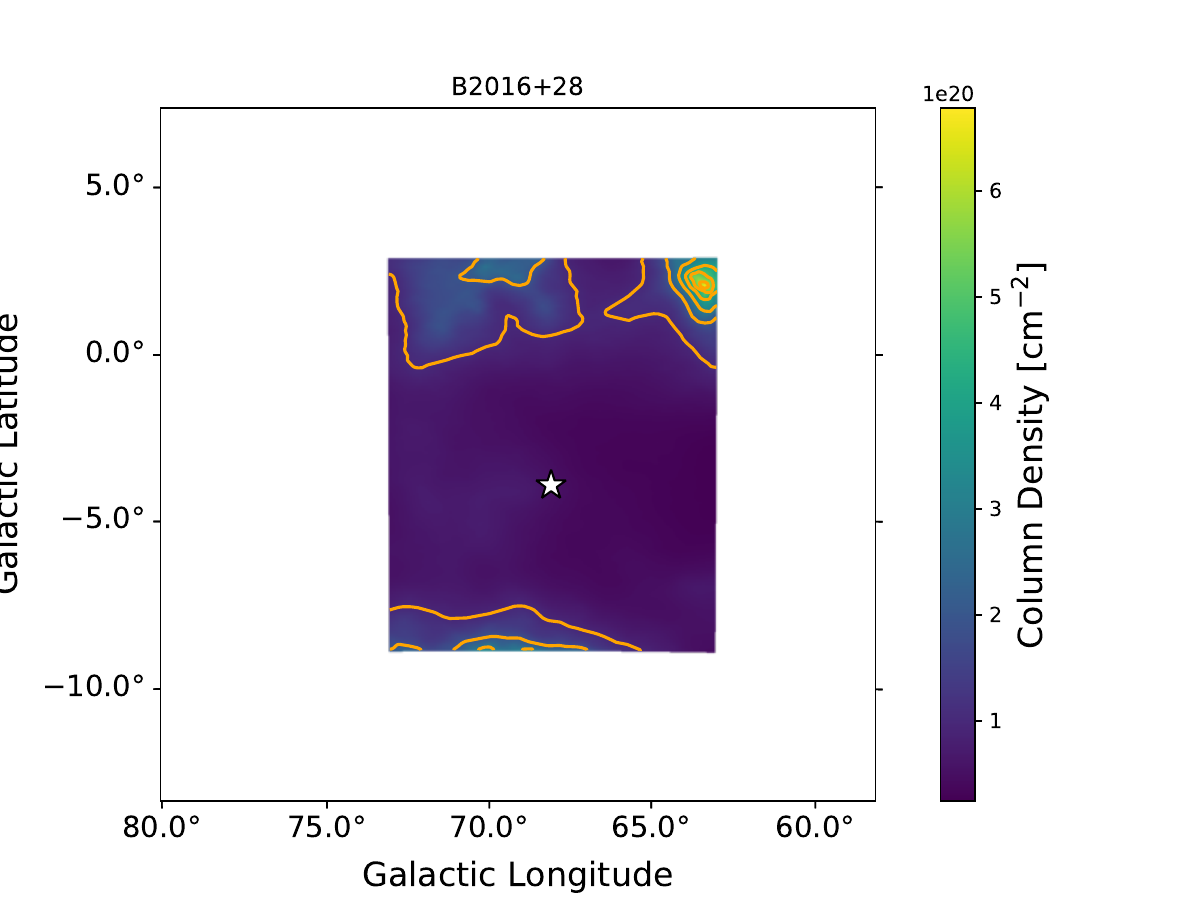}
    \includegraphics[width=0.40\textwidth]{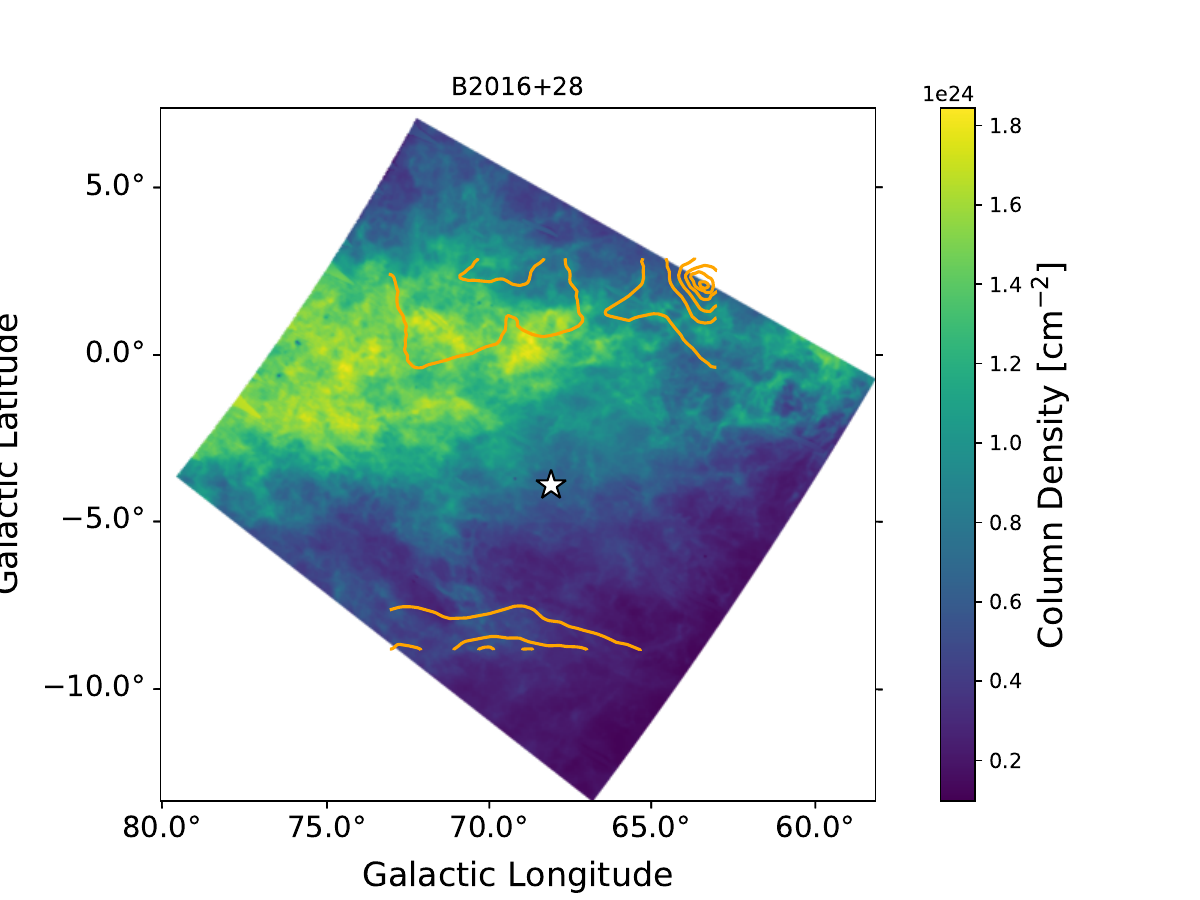}
    \caption{Total neutral hydrogen (left column) and \hi\ (right column) column density images of the auxiliary components toward B1929+10 and B2016+28. We integrate over the FWHM centered on the location of the density (velocity) peak for the total neutral hydrogen (\hi) column density images. The first two rows show the first and third component towards B1929+10, while subsequent rows show the first, second, fourth, and fifth components towards B2016+28, respectively. Pink contours denote six linearly spaced levels between the minimum and peak total neutral hydrogen column density. The pulsar position is marked by a white star.}
    \label{fig:appendix_integrated-cdensity}
\end{figure*}

\begin{figure*}
\ContinuedFloat
    \centering
    \includegraphics[width=0.40\textwidth]{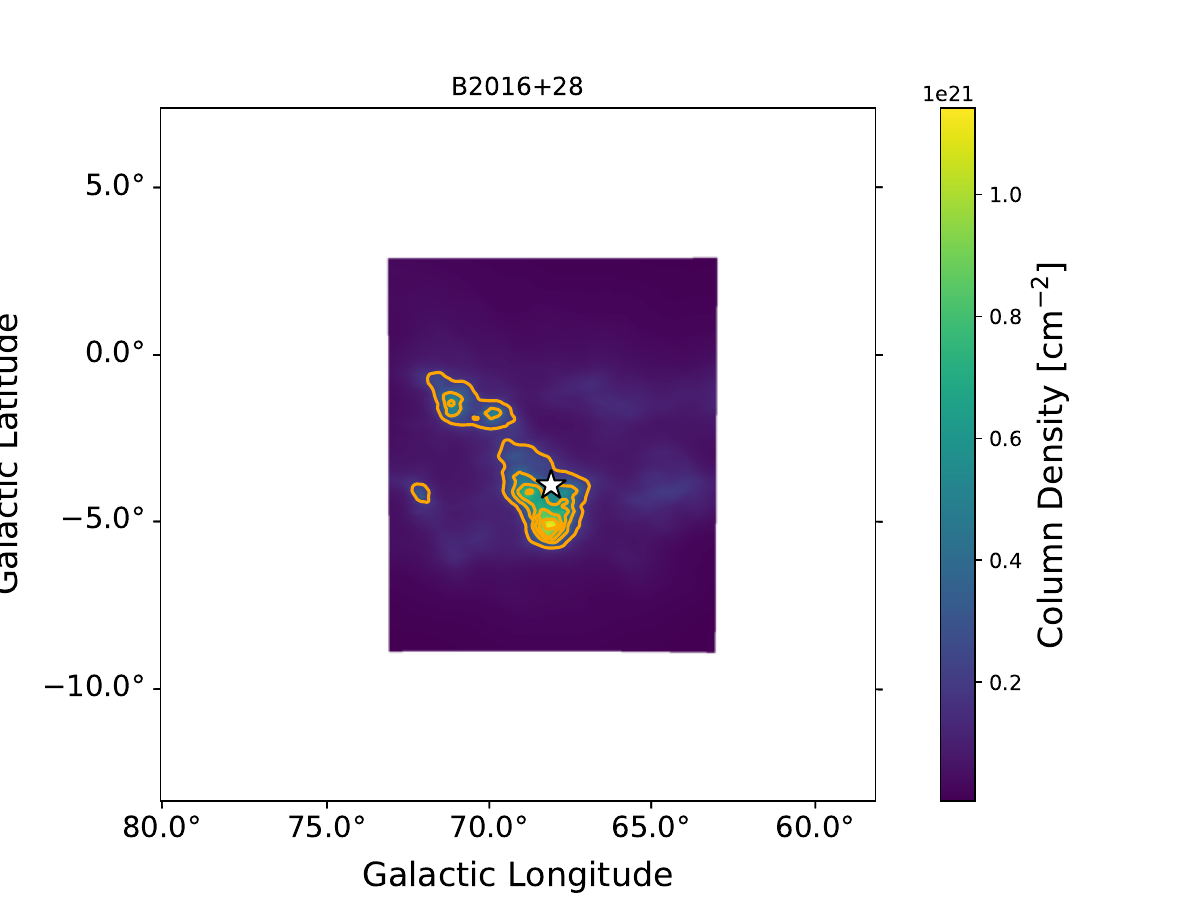}
    \includegraphics[width=0.40\textwidth]{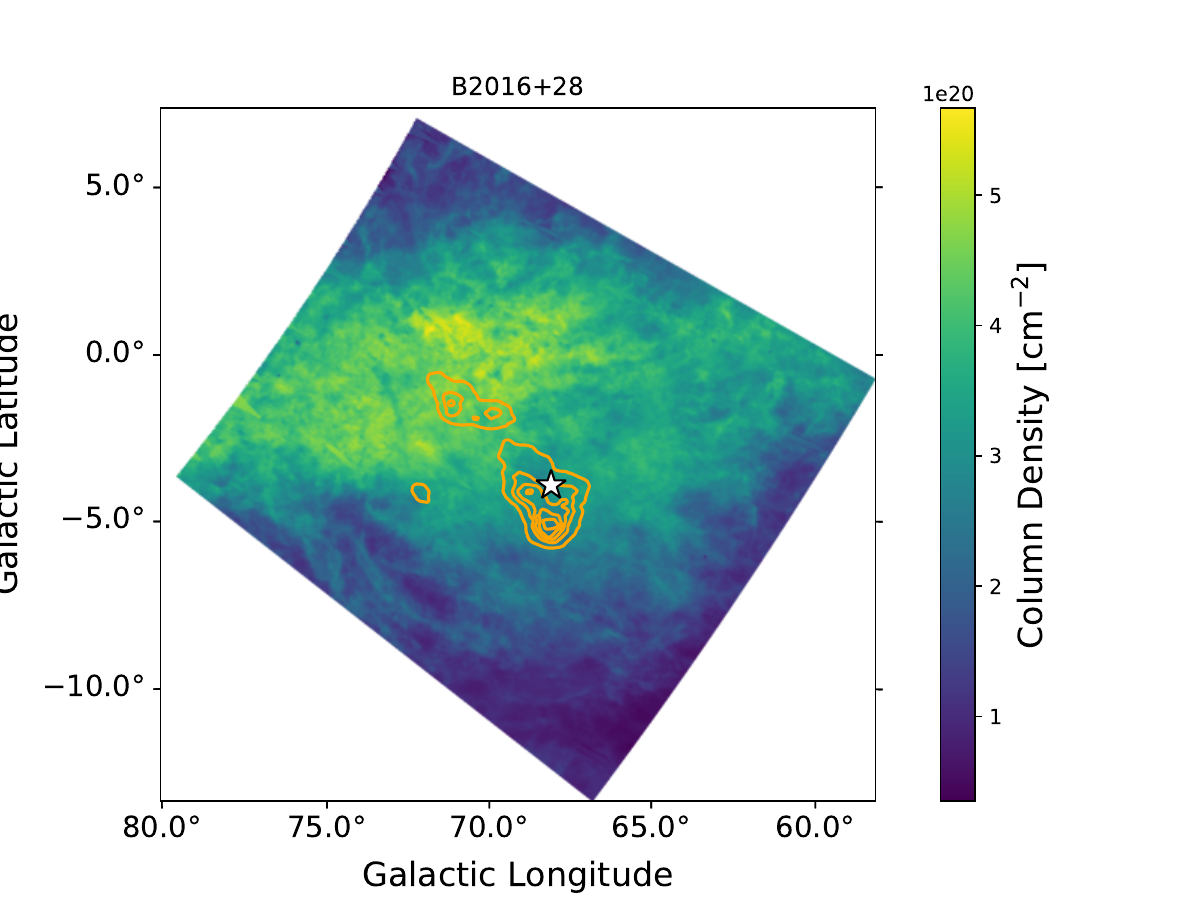}
    \includegraphics[width=0.40\textwidth]{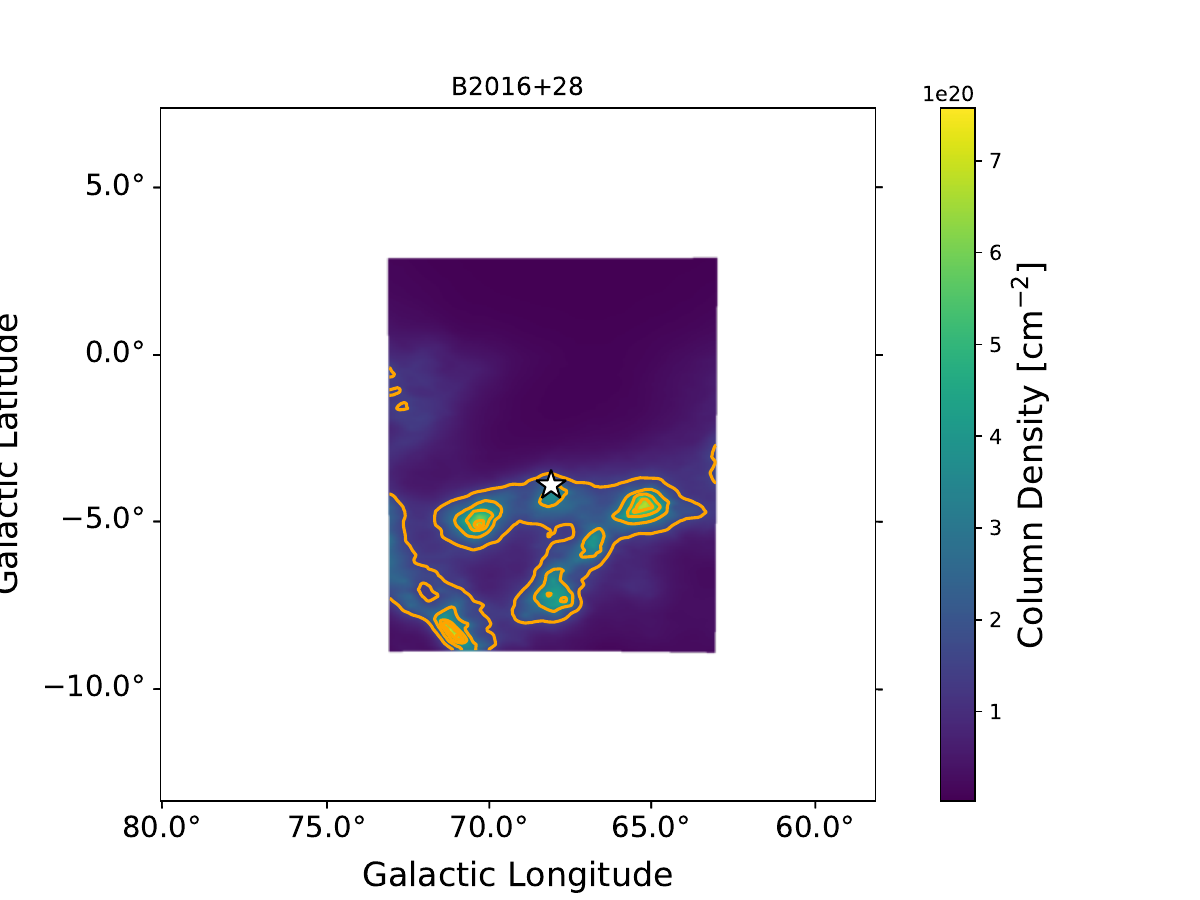}
    \includegraphics[width=0.40\textwidth]{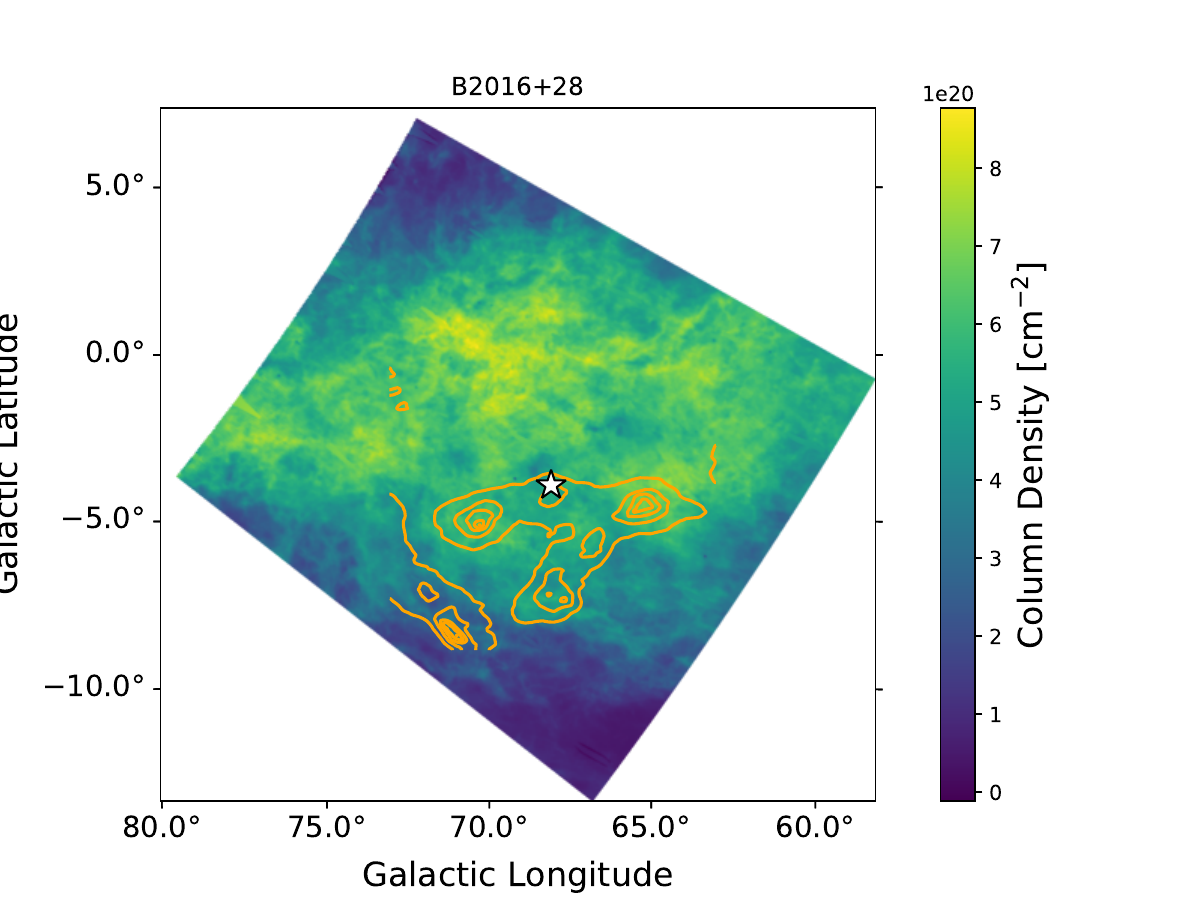}
    \includegraphics[width=0.40\textwidth]{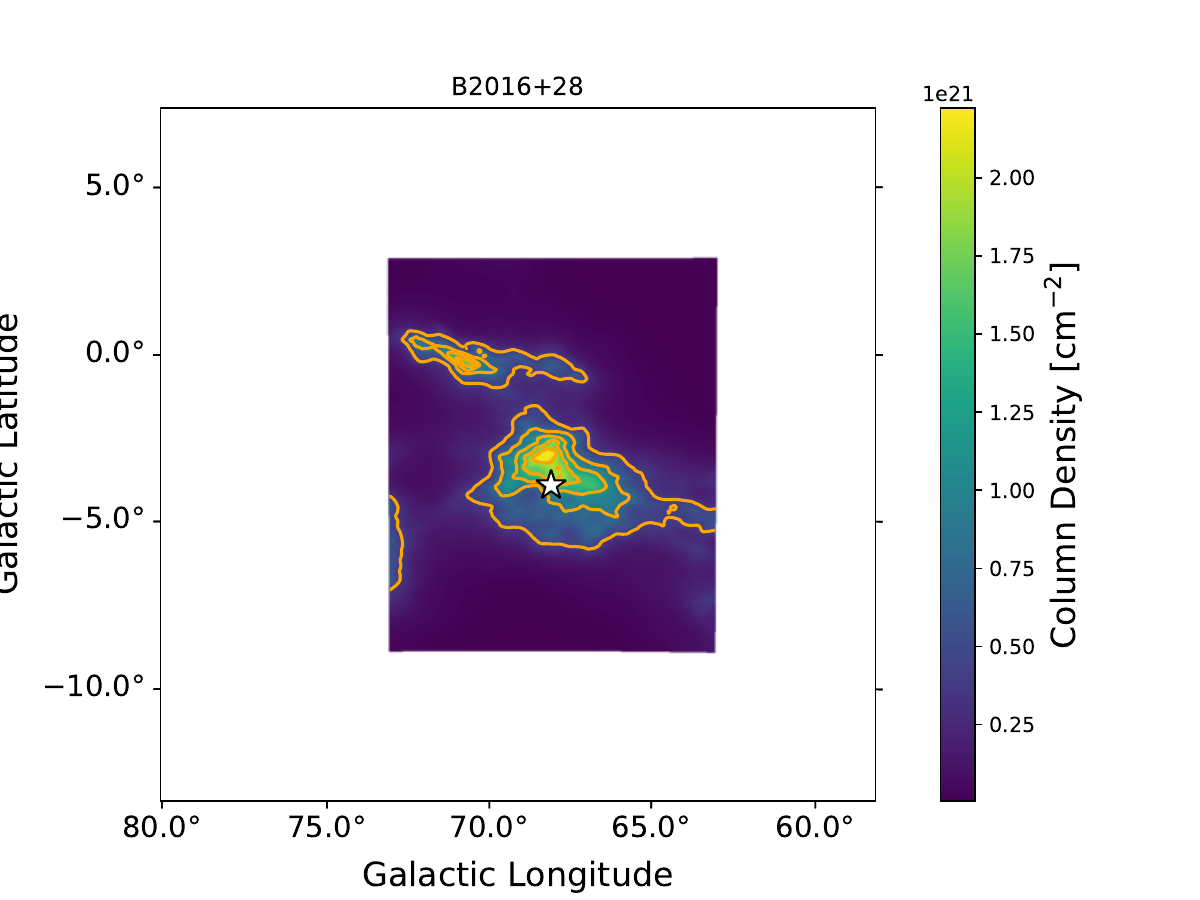}
    \includegraphics[width=0.40\textwidth]{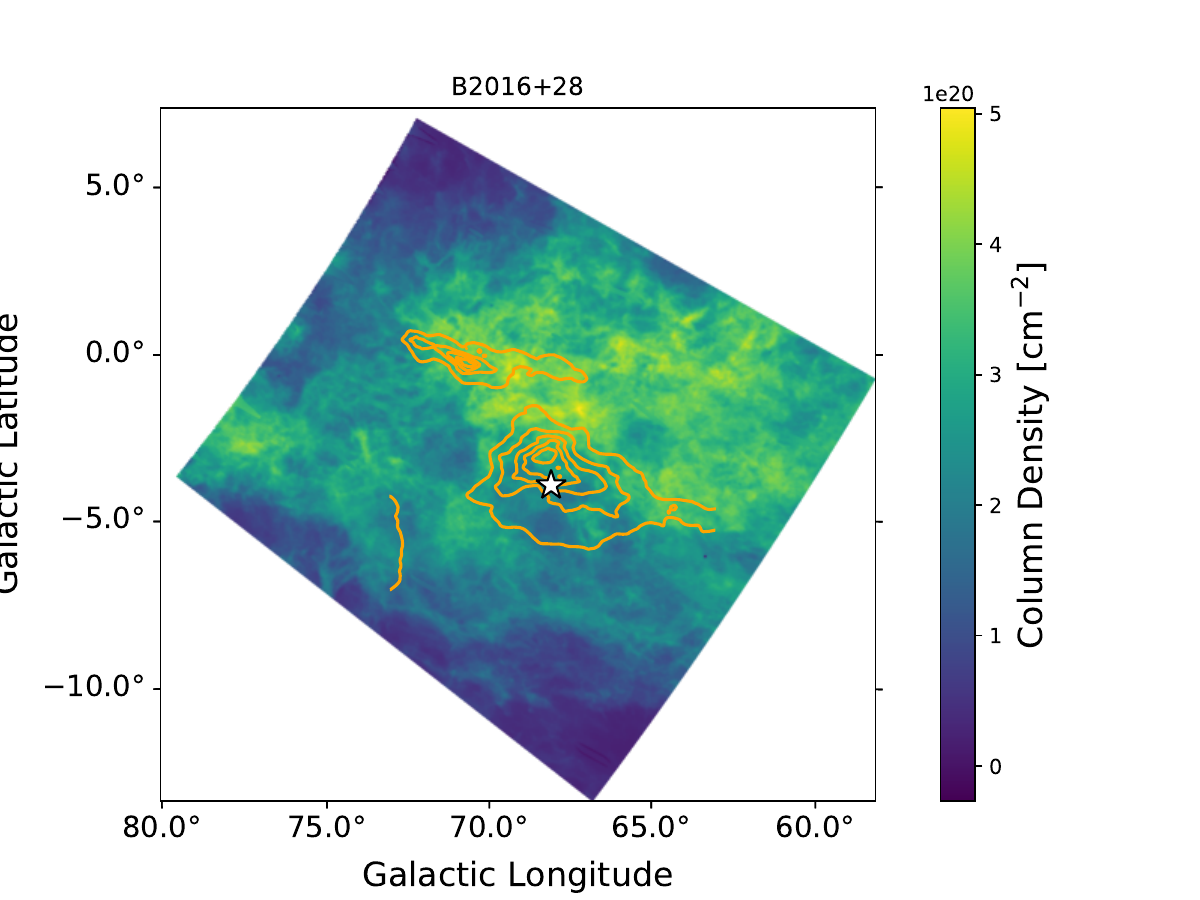}
    \caption[]{Figure~\ref{fig:appendix_integrated-cdensity} continued.}
\end{figure*}

\bibliographystyle{pasa-mnras}
\bibliography{all_refs}

\end{document}